\documentclass{elsarticle}

\usepackage[utf8]{inputenc}
\usepackage[T1]{fontenc}
\usepackage{graphicx} %
\usepackage{caption}
\usepackage{enumitem}
\usepackage{contour}
\usepackage{algorithm}
\usepackage{algorithmicx}
\usepackage[noend]{algpseudocode}
\usepackage{subcaption}
\usepackage{appendix}
\usepackage{anysize}
\usepackage{mathtools}
\usepackage{multicol}
\usepackage{breqn} 
\usepackage{soul}
\usepackage{nameref}
\usepackage{placeins}
\usepackage{xcolor}
\usepackage{colortbl}
\usepackage{ulem}
\usepackage{smartdiagram}
\usesmartdiagramlibrary{additions} %

\usepackage[rightcaption]{sidecap}

\usepackage{nomencl}
\usepackage{wrapfig}
\usepackage{comment}
\usepackage{pgfplots}
\usepackage[version=4]{mhchem}
\usepackage{amsmath}
\usepackage{dsfont} 
\usepackage{amsthm}
\usepackage{cancel}
\usepackage{ulem}
\usepackage{graphicx}
\usepackage{booktabs}
\usepackage{xspace}
\usepackage{nicematrix}
\usepackage{scalerel,stackengine}
\usepackage[colorlinks=true, allcolors=blue]{hyperref}
\usepackage{flowchart}
\usepackage{tikz}
\usepackage{tikz-3dplot}
\usetikzlibrary{shapes.geometric, arrows.meta, positioning}

\usetikzlibrary{fit}
\usetikzlibrary{positioning}
\usetikzlibrary{calc}
\usepackage{cleveref}

\newcommand{\mathbb}[1]{\mathds{#1}}

\newcommand{\liq}{l}
\newcommand{\gas}{g}
\newcommand{\phaseFrac}{\alpha^{\gas}} 
\newcommand{\liqFrac}{\alpha^{\liq}} 
\newcommand{\moleFrac}{\boldsymbol{z}}
\newcommand{\total}{\scriptscriptstyle\text{tot}}
\newcommand{\inp}{\scriptscriptstyle\text{in}}
\newcommand{\out}{\scriptscriptstyle\text{out}}
\newcommand{\TAlphaRho}{\scriptscriptstyle{T\alpha \rho}}
\newcommand{\Q}{S^{\scriptscriptstyle{T\alpha \rho}}}

\newcommand{\bulk}{\scriptscriptstyle\beta}
\newcommand{\incp}{\varepsilon}
\newcommand{\singlePhase}{\star}
\newcommand{\hllcF}{F}
\newcommand{\discreteU}{\mathbf{U}}
\newcommand{\discreteV}{\mathbf{V}}
\newcommand{\discreteF}{\mathbf{f}}
\newcommand{\discreteFPipe}{\mathbf{f}_{\scriptscriptstyle\text{Pipe}}}
\newcommand{\discreteFPipeI}[1]{\mathbf{f}_{\scriptscriptstyle\text{Pipe, #1}}}
\newcommand{\discreteFTank}{\mathbf{f}_{\scriptscriptstyle\text{Tank}}}

\newcommand{\coo}{\ensuremath{\mathrm{CO_2}}\xspace}
\newcommand{\nitrogen}{\ensuremath{\mathrm{N_2}}\xspace}
\newcommand{\methane}{\ensuremath{\mathrm{CH_4}}\xspace}

\newcommand{\oxygen}{\ensuremath{\mathrm{O_2}}\xspace}
\newcommand{\hydrogen}{\ensuremath{\mathrm{H_2}}\xspace}

\newcommand{\etal}{et al.}

\newcommand{\NExpanded}[1]{ N_1^{#1}, \dots, N_n^{#1}}
\newcommand{\NBold}[1]{\mathbf{N}^{#1}}
\newcommand{\NExpandedNoSuper}{ N_1, \dots, N_n}
\newcommand{\NBoldNoSuper}{\mathbf{N}} 
\newcommand{\Res}{\xi}

\newcommand{\cU}{{\mathcal{U}}}
\newcommand{\EulerFlux}{{\mathcal{F}}}
\newcommand{\Source}{\mathcal{S}}

\newcommand{\tvnSpaceArgs}{T, V, \NBoldNoSuper}

\newcommand{\rd}{\mathrm{d}}
\newcommand{\pd}[2]{\frac{\partial #1}{\partial #2}}
\newcommand{\pdBracket}[3]{\left(\pd{#1}{#2}\right)_{#3}}

\newcommand{\dd}[2]{\frac{\rd #1}{\rd #2}}

\allowdisplaybreaks
\stackMath

\title{A Reformulation of UVN-Flash for Multicomponent Two-Phase Systems with Application to \coo-rich Mixture Transport in Pipelines}

\makenomenclature
\begin{document}

\begin{frontmatter}

\author[delft,cwi]{Pardeep Kumar\corref{cor1}}
\ead{pardeep@cwi.nl}

\author[shell]{Patricio I. Rosen Esquivel}

\cortext[cor1]{Corresponding author}

\affiliation[delft]{
    organization={Delft University of Technology},
    city={Delft},
    country={The Netherlands}
}

\affiliation[cwi]{
    organization={Centrum Wiskunde \& Informatica},
    city={Amsterdam},
    country={The Netherlands}
}

\affiliation[shell]{
    organization={Shell Projects and Technology},
    city={Amsterdam},
    country={The Netherlands}
}

\begin{abstract}
Pipeline transport of dense-phase \ce{CO2}--rich mixtures is a crucial component in carbon capture and storage (CCS). Accurate modeling requires coupling of fluid dynamics and thermodynamics, especially during transient events such as depressurization. In this work, we present a unified framework for two-phase multicomponent transport in pipelines that integrates both aspects. Specifically, we employ the homogeneous equilibrium model (HEM) for modeling the transport of two-phase \ce{CO2}--rich mixture, with thermodynamic closure provided by a Helmholtz energy-based equation of state. Phase equilibrium calculations are performed using UVN-flash, supplemented with a stability analysis procedure to detect phase separation and generate initial guesses for the phase-equilibrium calculations. Specifically, we introduce a novel tailored UVN-flash routine that aligns with the fluid dynamics formulation. This is achieved by introducing an alternative and better-scaled set of variables for the phase-equilibrium calculations. The proposed framework is applied to the depressurization of tanks and pipelines containing \ce{CO2}--rich mixtures, demonstrating its effectiveness for CCS--relevant applications.

\end{abstract}


\end{frontmatter}

\section{Introduction}
 According to the IEA~\cite{noauthor_iea_2023}, carbon capture and storage (CCS) is expected to play an important role in addressing the challenge of reducing greenhouse gas emissions. In CCS, \ce{CO2} is captured from industrial sources and transported to storage sites, e.g., depleted gas fields, where it is injected underground for permanent storage. A major share of this transport is expected to be carried out via pipelines. The captured \ce{CO2} from industries is generally not pure and can contain varying levels of impurities such as methane, nitrogen, amines, sulfur oxides, and water etc.~\cite{munkejord_thermo-_2010}. These impurities can significantly affect phase behavior, thermophysical properties, and consequently, the design and safe operation of pipeline systems.

One of the extreme events in pipeline operations is depressurization, which can be required following a shut-in operation~\cite{munkejord_depressurization_2015, munkejord_depressurization_2021, log_temperature_2025}. In such events, the temperatures and pressures can reach extreme values with the risk of exceeding the material operational limit of the pipe wall. Accurately predicting this behaviour is essential for safe pipeline design.  In this paper, we investigate the depressurization from a pipeline carrying \ce{CO2} with impurities.

Some experimental studies have investigated decompression behaviour of \ce{CO2}--rich mixtures, providing valuable reference data for model validation.
Drescher \etal~\cite{drescher_experiments_2014} provided experimental data for \ce{CO2-N2} mixtures. Cosham \etal~\cite{cosham_decompression_2012} presented 14 shock-tube experiments, including tests with pure \ce{CO2} as well as \ce{CO2} containing impurities, to study decompression behaviour. Botros \etal~\cite{botros_measuring_2013} reported experimental data for \ce{CO2-CH4} mixtures to determine decompression wave speeds. 
 
Two-phase flow of fluid mixtures is described by the Navier–Stokes equations. For general two-phase flow models, the reader is referred to Stewart \etal~\cite{bruce_stewart_two-phase_1984}, and for \ce{CO2} transport specifically, to the review by Munkejord et al.~\cite{munkejord_co2_2016} and the references therein. For pipeline applications, cross-sectional averaging is typically applied to obtain one-dimensional two-fluid models~\cite{toumi_approximate_1999}. Such averaging may render the system of equations ill-posed. To avoid these issues and focus more on the integration between the fluid dynamics and thermodynamics, we employ the Homogeneous Equilibrium Model (HEM), which is hyperbolic and guarantees well-posedness. In HEM, the fluid is treated as a homogeneous mixture of two well-mixed phases, with the Navier-Stokes equations formulated directly in terms of mixture density, velocity, and energy. For an inviscid flow in a frictionless pipe, this formulation reduces to Euler's equations, where all conserved quantities are expressed in terms of mixture properties.  

In addition to the fluid dynamics, the phase behaviour of the real fluids plays a crucial role and is governed by thermodynamic principles. Depending on the operating conditions, \ce{CO2} transport in pipelines can either be single-phase or two-phase. To describe two-phase flow accurately, a real-gas equation of state (EOS) is required. Often, the Mie-Grüneisen stiffened gas EOS is used; see~\cite{saurel_multiphase_1999, saurel_relaxation-projection_2007, abgrall_computations_2001, kapila_two-phase_2001, pelanti_mixture-energy-consistent_2014}. However, to accurately describe the real fluid behavior, more advanced EOSs are required. These advanced EOSs are defined in terms of the Helmholtz free energy function: given the temperature, volume, and composition of a single-phase, they allow calculation of all thermodynamic properties.  It is important to note that such EOSs are defined at the single-phase level; thermodynamic properties cannot be directly evaluated from overall mixture quantities in a two-phase system without first performing a phase-split calculation.
 In a fluid dynamic simulation, the mixture properties, namely the volume (mesh size), velocity, and energy, are known at each time step. From these quantities, one intends to compute the mixture pressure. Naturally, this leads to a phase-split calculation, commonly referred to in thermodynamic literature as a \textit{flash calculation}.

 A substantial body of literature exists on flash calculations. The majority focuses on the PTN-flash (pressure--temperature specified) problem, (see~\cite{michelsen_isothermal_1981, michelsen_isothermal_1982, michelsen_thermodynamic_2007, holyst_thermodynamics_2012}. The second most documented case is the VTN-flash (volume--temperature specified), (see~\cite{nichita_calculation_2007, nichita_isochoric_2009, nichita_rapid_2013, nichita_robustness_2023}). Both PTN and VTN are isothermal flash problems. The literature, however, is rather scarce on the non-isothermal flashes, for instance, the PHN-flash (pressure--enthalpy specified), the UVN-flash (internal energy--volume specified), or the SVN-flash (entropy--volume specified). For fluid dynamical simulations, UVN-flash provides a natural formulation, since the internal energy, specific volume, and molar composition are available in each computational cell. Before carrying out the flash calculation, it is essential to know whether the fluid mixture is in a single-phase or two-phase condition. This is determined via stability analysis. If stability analysis predicts the existence of a two-phase state, the flash calculation is performed. Moreover, the outcome of the stability analysis often provides an excellent initial guess for the subsequent flash calculation.

Some researchers have investigated the numerical simulation of \ce{CO2}--transport in pipelines.  Elshahomi \etal~\cite{elshahomi_decompression_2015} presented numerical simulations of decompression-wave speeds using ANSYS software. Over the past several years, SINTEF Energy Research in Norway has been a leading contributor in this area. They have developed a dedicated \ce{CO2} flow loop for experiments and have published a steady stream of work on \ce{CO2} transport;  see~\cite{ munkejord_depressurization_2015, munkejord_co2_2016, munkejord_depressurization_2020}. Their work covers both numerical modeling of \ce{CO2} transport with and without impurities and detailed studies of thermodynamic behaviour. Most studies focus either on the fluid dynamics; see~\cite{ munkejord_depressurization_2015, munkejord_depressurization_2020, hammer_method_2013, morin_two-fluid_2013} and use an existing thermodynamic library as-is, or they focus on the thermodynamic behavior~\cite{aursand_spinodal_2017, lachet_equilibrium_2012, wilhelmsen_thermodynamic_2017}.  

In this paper, we present a comprehensive treatment of two-phase multicomponent transport of fluid mixtures in pipelines that addresses both the fluid dynamics aspects as well as the thermodynamic aspects with a UVN-flash framework. In particular, we focus on the robustness of the flash problem for challenging inputs from the fluid dynamics solver. To address these challenges, we propose a reformulation of the UVN-flash problem through an alternative set of variables. This transformation improves the robustness of phase-equilibrium calculations. The key contribution of this paper is the development and testing of the flash routine tailored for dynamic pipeline transport models. The proposed methodology is tested using both tank and pipeline depressurization cases. 

The paper is structured as follows. We begin with the governing equations for HEM model and the dynamics of Tank depressurization in \Cref{sec:dynamics}. In \Cref{sec:thermodynamics}, we discuss the thermodynamic aspects including the calculation of the thermodynamic properties and the choice of the EOS. Stability analysis and flash calculations for phase-split problems are discussed in \Cref{app:stability_analysis} and \ref{sec:flash}, respectively. The time discretization method is addressed in \Cref{sec:time_discretization}. Finally, \Cref{sec:results} presents the results and compares them with literature data, followed by the conclusions in \Cref{sec:conclusion}.
\section{Governing Equations} \label{sec:dynamics}

This section describes the governing equations for the fluid flow.

\subsection{Fluid flow equations in pipeline} \label{subsec:HEM}
The Navier-Stokes(NS) equations describe the dynamics of fluid flow. In principle, one could solve the 3D equations; however, for practical applications where a pipeline could be many kilometers long, it becomes computationally prohibitive. For these situations, a more efficient approach is to use a cross-sectionally averaged 1D equations~\cite{ishii_thermo-fluid_2011}. For a fully dispersed two-phase fluid flow, the governing equations can be written in terms of average mixture quantities, leading to the Homogeneous Equilibrium Model (HEM). For inviscid flow through a horizontal frictionless pipe, neglecting heat transfer effects, the HEM can be expressed as:

\begin{equation}
\label{Eqns:NS}
\partial_{t} \cU + \partial_{x} \EulerFlux(\cU) = \Source(\cU),
\end{equation}
where $\cU(x,t)$ is the vector of conserved variables, $\EulerFlux(\cU)$ the flux vector and $\Source(\cU)$ the source terms. These terms have the following form:
\begin{equation}
\label{vector_eqns}
\cU = \begin{bmatrix}
{\rho }\\
{\rho u }\\
{\rho E}
\end{bmatrix}, \quad
\EulerFlux(\cU) = \begin{bmatrix}
{\rho u}\\
{\rho u^{2} + p }\\
{(\rho E + p) u}
\end{bmatrix} \quad
\Source(\cU) = \begin{bmatrix}
{0}\\
{0}\\
{0}
\end{bmatrix}.
\end{equation}
Here, \(\rho, u, E, p\) denote the mixture mass density, mixture velocity, specific total energy, and the pressure, respectively. The HEM model assumes instantaneous thermodynamic equilibrium and that both phases travel at the same speed. The total specific energy $E$ can be expressed as
\begin{align}
    E = e + \frac{1}{2} u^2,
\end{align}
where \(e\) is the specific internal energy. The mixture properties can be written in terms of the phasic properties as follows
\begin{subequations}
\begin{align}    
    \rho   &:= \alpha \rho_{\gas} + (1 - \alpha) \rho_{\liq}, \label{Def:density} \\
    \rho e &:= \alpha \rho_{\gas} e_{\gas}+ (1 - \alpha) \rho_{\liq} e_{\liq}, \label{Def:Energy}
\end{align}    
\end{subequations}
where \(\alpha\) denotes the volume fraction of the gas phase, \(\rho_{\gas}\) and \(\rho_{\liq}\) denote the mass density of the gas and liquid phase, respectively, \(e_{\gas}\) and \(e_{\liq}\) denote the specific internal energy (mass basis) of the corresponding phases.

The system of equations \eqref{Eqns:NS} is closed by incorporating an equation of state(EOS), which relates pressure to the other thermodynamic quantities. The role of the EOS in closing the system will be discussed in detail in \Cref{sec:thermodynamics}. Furthermore, for the problem to be well-posed, appropriate initial and boundary conditions must be specified; these aspects will be discussed in \Cref{sec:results}.

\subsection{Spatial Discretization} \label{subsec:spatial_discretization}
The HEM model described in the \Cref{subsec:HEM} is solved using the Finite Volume Method (FVM). As a first step, we discretize the system \eqref{Eqns:NS} in space, which yields the following semi-discrete form:
\begin{align} \label{eqn:semiDiscrete}
\dd{\discreteU_i}{t} = -\frac{1}{\triangle x} (\Hat{\EulerFlux}_{i+\frac{1}{2}} - \Hat{\EulerFlux}_{i-\frac{1}{2}}), \qquad i=1\ldots N,
\end{align}
where $\discreteU_i(t) \in \mathbb{R}^{3} \approx \cU(x_{i},t)$, represents the cell-averaged conserved variables in cell \(i\), ${\triangle x}$ is the spatial grid spacing, $\Hat{\EulerFlux}_{i\pm\frac{1}{2}}$ denote the numerical fluxes  at the cell interfaces \(i\pm\frac{1}{2}\) and $N$ is the total number of finite volume cells. System \eqref{eqn:semiDiscrete} for the full computational domain can be expressed compactly in vectorial notation as:
\begin{equation}\label{eq:pipe:vector_form}
    \frac{\rd \discreteU(t)}{\rd t} = \discreteFPipe(\discreteU(t), \discreteV(t)),
\end{equation}
where \(\discreteU(t) \in \mathbb{R}^{3N}\) is the global vector of conserved quantities for the entire domain(pipe), defined as:
\begin{align*}
  \discreteU(t) = [\discreteU_{1}(t), \ldots, \discreteU_{i}(t), \ldots, \discreteU_{N}(t)]^T, \qquad \discreteU_{i} = [\rho_i, (\rho u)_i, (\rho E)_i]^T,
\end{align*}
Here, \(\discreteU(t)\) is composed of $N$ individual cell vectors \(\discreteU_i(t)\), and \(\rho_i, (\rho u)_i\) and \((\rho E)_i\) denote the mass density, momentum, and the total energy in the \(i^{th}\) cell.
The vector \(\discreteV(t)\) contains the non-conservative (algebraic) variables associated with the thermodynamic state in each cell. Its precise definition will be provided in ~\Cref{sec:time_discretization}.
\newline
The right-hand side \(\discreteFPipe\) depends on both the conservative variables \(\discreteU(t)\) and the associated non-conservative variables \(\discreteV(t)\), and is defined as
\begin{align}
  \discreteFPipe(\discreteU(t), \discreteV(t)) := [\discreteFPipeI{1}(t), \ldots, \discreteFPipeI{i}(t), \ldots, \discreteFPipeI{N}(t)]^T,
\end{align}
where each local \(\discreteFPipeI{i}\) is expressed as
\begin{align}
  \discreteFPipeI{i}(\discreteU_{i-r:i+r}, \discreteV_{i-r:i+r}) = -\frac{1}{\Delta x} \left( \Hat{\EulerFlux}_{i+\frac{1}{2}} - \Hat{\EulerFlux}_{i-\frac{1}{2}} \right).
\end{align}
where \(r\) denotes the stencil radius (half-width), \(\discreteU_{i-r:i+r} := (\discreteU_{i-r}, \ldots, \discreteU_{i+r})\), and similarly for \(\discreteV_{i-r:i+r}\). The numerical fluxes \(\Hat{\EulerFlux}_{i\pm\frac{1}{2}}\) are computed using data from a stencil of width \(2r + 1\), and hence depend on the conservative variables \(\discreteU_j\) and the corresponding non-conservative variables \(\discreteV_j\) for \(j = i - r, \ldots, i + r\). For a first-order scheme, we put \(r = 1\). To compute the inter-cell numerical fluxes $\Hat{\EulerFlux}_{i\pm\frac{1}{2}}$, we employ the HLLC scheme; (see \ref{HEM_HLLC}).

\subsection{Tank depressurization} \label{gov_eqns.sec.tank}

 The tank model can be regarded as a pipeline discretised with a single computational cell. Many researchers have previously considered the dynamic simulation of the tank in the context of UVN-flash\cite{castier_dynamic_2010, arendsen_dynamic_2009, saha_isoenergetic-isochoric_1997, lima_differential-algebraic_2008}. Furthermore, this model has also been employed by \cite{qiu_multiphase_2014, giljarhus_solution_2012, kumar_new_2025} in the context of pipeflows. The governing equations describing the evolution of the molar composition and internal energy of the tank are given by:

\begin{subequations}
    \begin{align}
        &\dd{N_i}{t} = %
        \dot{N}_{i}^{\inp} - %
        \dot{N}_{i}^{\out} \quad \text{for } i = 1,\dots,n \label{tank.evolution.moles}, \\ 
        &\dd{U}{t} = %
        \dot{H}^{\inp} - %
        \dot{H}^{\out} + \dot{Q}, \label{tank.evolution.int_energy}
    \end{align}
\end{subequations}
where \(n\) denotes the total number of components in the mixture. The variable \(N_{i}\) represents the moles of component \(i\) in the tank, while \(U\) is the total internal energy. The terms \(\dot{N}_{i}^{\inp}\) and \(\dot{N}_{i}^{\out}\) denote the molar flow rates of component \(i\) in the input and output streams, respectively. Likewise, \(\dot{H}^{\inp}\) and \(\dot{H}^{\out}\) denote the enthalpy flow rates associated with the input and output streams, and \(\dot{Q}\) is the rate of heat supplied to the tank.

The equations \eqref{tank.evolution.moles}–-\eqref{tank.evolution.int_energy} can be written in compact vector form as:
\begin{equation} \label{eq:tank:vector_form}
    \frac{\mathrm{d} \discreteU}{\mathrm{d} t} = \discreteFTank(\discreteU(t), \discreteV(t)),
\end{equation}
where the state vector \(\discreteU(t) \in \mathbb{R}^{n+1}\) is defined as
\begin{align}
    \mathbf{\discreteU}(t) = \begin{bmatrix}
            N_1(t) \\
            \vdots \\
            N_n(t) \\
            U(t)
        \end{bmatrix},
\end{align}
and the vector \(\discreteV(t)\) contains the variables associated with the thermodynamic state. This will be revisited in \Cref{sec:time_discretization}. The right-hand side \(\discreteFTank(\discreteU, \discreteV)\) is given by
\begin{align} \label{eq:tank_rhs}
    \discreteFTank(\discreteU, \discreteV) =
    \begin{bmatrix}
        \displaystyle %
        \dot{N}_{1}^{\mathrm{in}} - %
        \dot{N}_{1}^{\mathrm{out}} \\
        \vdots \\
        \displaystyle %
        \dot{N}_{n}^{\mathrm{in}} - %
        \dot{N}_{n}^{\mathrm{out}} \\
        \displaystyle %
        \dot{H}^{\mathrm{in}} - %
        \dot{H}^{\mathrm{out}} + \dot{Q}
    \end{bmatrix}.
\end{align}

\section{Thermodynamical aspects} \label{sec:thermodynamics}
\subsection{Definitions}
For the sake of clarity, we define the following thermodynamic constructs under specified constraints of total internal energy $(U)$, volume $(V)$, and mole numbers \(\NBoldNoSuper\).
\newline\newline
\textbf{Reference Phase:} The reference phase (\(\singlePhase\)) represents a hypothetical single-phase state characterized by the total internal energy \( U^{\singlePhase} \), volume \( V^{\singlePhase} \) and total mole numbers \( \NBold{\singlePhase} = (\NExpanded{\singlePhase}) \). 
\newline\newline
\textbf{Trial Phase:} The trial phase is an incipient(very small amount) phase introduced to assess the thermodynamic stability of a system. The system is perturbed by adding this phase and if the total entropy increases, the system is considered unstable as a single phase, and phase separation is favorable. 
\newline\newline
\textbf{Stability Analysis:} 
Stability analysis determines whether a given fluid at specified \(U, V\) and \(\NBoldNoSuper\) will remain stable as a single phase or separate into two phases, i.e., into a gas-liquid mixture. 
If the single-phase state is unstable, this analysis provides estimates of its temperature, molar concentration (moles per unit volume), and molar internal energy density. These estimates are used to obtain initial guesses for the subsequent flash calculation. Please see the \ref{app:stability_analysis} for more details.
\newline\newline
\textbf{Flash:} 
A flash calculation determines the equilibrium phase split of a multicomponent mixture. Under specified thermodynamic constraints (e.g., UVN), it computes the distribution of extensive quantities among the coexisting phases, i.e., the enthalpy, internal energy, entropy, and volume of each phase, and the common intensive quantities, e.g., temperature, pressure, and component-wise chemical potential, which satisfy the conditions of thermal, mechanical, and chemical equilibrium.

\subsection{Equation of State}
To close the HEM model and the tank model, we need a constitutive relation to compute the pressure in the pipe model and the enthalpy flow rate in the tank model. This constitutive model is based on the thermodynamic properties of the fluid mixture and is commonly referred to as the Equation of State(EOS). The EOS for the fluid mixture is specified in terms of any two thermodynamic variables, such as pressure and temperature or volume and temperature, etc., in addition to the mole numbers of the components in the mixture. Thus, for an \(n\)-component mixture, \(n + 2\) variables are required to fully specify the thermodynamic state of the system. Based on the choice of the two input thermodynamic variables, the EOS can be classified into two categories: \textit{pressure-based} and \textit{volume-based}. Typically, the real fluid EOS is defined in terms of the Helmholtz free energy \(A\) as a function of volume \(V\), temperature \(T\), and the molar composition vector \(\NBoldNoSuper\): 
\begin{equation}
    A = A(\tvnSpaceArgs), \label{eqn:helmholtz_free_energy}
\end{equation}
where \(\NBoldNoSuper := \{\NExpandedNoSuper\}\) and \(N_i\) denote the number of moles of the \(i^{th}\) component in the fluid mixture. We use Peng--Robinson EOS~\cite{peng_new_1976} in this work. The various thermodynamic properties can be computed directly from the Helmholtz energy as below.

\begin{subequations} 
\begin{align}
    p &= -\pdBracket{A}{V}{T, \NBoldNoSuper} \\
    S &= -\pdBracket{A}{T}{V, \NBoldNoSuper} \\
    U &= A + T S \\
    H &= A + T S + p V \\
    C_v &= \pdBracket{U}{T}{V, \NBoldNoSuper} \\
    \mu_i &= \pdBracket{A}{N_i}{V, \{N_j\}_{j \ne i}}
\end{align}    
\end{subequations}
where \(p, S, U, H, C_v\) and \(\mu_i\) denote the pressure, entropy, internal energy, enthalpy, heat capacity at constant volume, and the chemical potential of the \(i^{th}\) component. Here, it is understood that all derivatives and functions are evaluated as functions of \(T\), \(V\), and \(\NBoldNoSuper\). 

An EOS is defined for individual homogeneous phases and must be applied to phasic variables, namely, the temperature, volume, and molar composition of each phase. It cannot be applied directly to mixture-averaged quantities in a multiphase system. However, in practical scenarios, only the extensive properties such as the total volume and the overall molar composition are known, and the distribution of the total volume and the mole numbers among the individual phases is not available. This necessitates a phase-split computation whereby one determines the distribution of the extensive quantities among the coexisting phases and the common intensive quantities (more on this later). 
The phase-split calculations are commonly referred to in the thermodynamic literature as the Flash calculations. 

\subsection{Choice of Flash}
Different types of flash problems arise depending upon which two thermodynamic variables are specified. The most common flash problem addressed in the literature is the PTN-Flash \cite{michelsen_isothermal_1981}, in which the pressure, temperature, and the overall molar composition are specified. The objective of the PTN-flash is to find the equilibrium phase split and compute the extensive properties of each phase, such as the internal energy, enthalpy, and volume etc. Another important flash problem is the VTN-Flash, where the total volume of the mixture is specified along with the temperature and overall composition, see for example~\cite{nichita_new_2018}. Both PTN and VTN-flashes are commonly referred to as isothermal flashes since the temperature is fixed.

In closed systems, for example, the energy balance calculation in process design, the relevant specification often includes the total internal energy \(U\), total volume \(V\), and overall molar composition \(\NBoldNoSuper\). This non-isothermal specification corresponds to the so-called UVN-flash problem. In our tank model and the HEM model, we typically have access to the total volume (fixed for the tank, cell volume for the pipe), total internal energy, and molar composition at each time step of the simulation. Thus, UVN-Flash formulation provides a natural approach to determine equilibrium phase-split and, consequently, the pressure in such scenarios.

\section{Stability Analysis} \label{app:stability_analysis}
In this section, we briefly review the stability analysis procedure. For a detailed account on this topic, we refer the interested reader to the existing literature \cite{michelsen_isothermal_1982, smejkal_phase_2017, nichita_robustness_2023, kumar_solving_2026}.  

\subsection{UVN stability}

To assess the thermodynamic stability of a single-phase system with fixed total internal energy \( U^\singlePhase \), volume \( V^\singlePhase \), and mole numbers \( \NBoldNoSuper^\singlePhase = (N_1^\singlePhase, \dots, N_n^\singlePhase) \), we consider a hypothetical phase separation into two phases: a bulk phase \(\beta\) and an incipient (infinitesimal) phase \(\incp\). The total entropy of the resulting two-phase system is given by:

\begin{align}
    S_{\text{two-phase}} = S(U^{\bulk}, V^{\bulk}, \NBoldNoSuper^{\bulk}) + S(U^{\incp}, V^{\incp}, \NBoldNoSuper^{\incp})
\end{align}
subject to conservation constraints:
\begin{align}
    U^\singlePhase &= U^{\bulk} + U^{\incp}, \\
    V^\singlePhase &= V^{\bulk} + V^{\incp}, \\
    N_i^\singlePhase &= N_i^{\bulk} + N_{i}^{\incp}, \qquad i=1 \dots n.
\end{align}
The entropy of the reference phase is:
\begin{align}
    S_{\text{ref}} = S(U^\singlePhase, V^\singlePhase, \NBoldNoSuper^\singlePhase).
\end{align}
The bulk phase entropy can be computed by performing a first-order Taylor expansion of the entropy around the reference state \((U^\singlePhase, V^\singlePhase, \NBoldNoSuper^\singlePhase)\), treating the incipient phase contribution as an infinitesimal perturbation:
\begin{align}
S^{\bulk} &:= S(U^\singlePhase - U^{\incp}, V^\singlePhase - V^{\incp}, N_1^\singlePhase - N_{1}^{\incp}, \dots, N_n^\singlePhase - N_{n}^{\incp}), \\
&= S(U^\singlePhase, V^\singlePhase, \NBoldNoSuper^\singlePhase)
- U^{\incp} \pdBracket{S}{U}{V, \NBoldNoSuper}
- V^{\incp} \pdBracket{S}{V}{U, \NBoldNoSuper}
- \sum_{i=1}^n N_{i}^{\incp} \pdBracket{S}{N_i}{U, V, \NBoldNoSuper}.
\end{align}
This leads to
\begin{align}
    S_{\text{two-phase}} = S(U^\singlePhase, V^\singlePhase, \NBoldNoSuper^\singlePhase)
- U^{\incp} \pdBracket{S}{U}{V, \NBoldNoSuper}
- V^{\incp} \pdBracket{S}{V}{U, \NBoldNoSuper}
- \sum_{i=1}^n N_{i}^{\incp} \pdBracket{S}{N_i}{U, V, \NBoldNoSuper} + S(U^{\incp}, V^{\incp}, \NBoldNoSuper^{\incp}).
\end{align}
The entropy difference \(\Delta S = S_{\text{two-phase}} - S_{\text{ref}}\) is then:
\begin{align} \label{eq:deltaS_1}
\Delta S  =
- U^{\incp} \pdBracket{S}{U}{V, \NBoldNoSuper}(U^\singlePhase, V^\singlePhase, \NBoldNoSuper^\singlePhase),
- V^{\incp} \pdBracket{S}{V}{U, \NBoldNoSuper}(U^\singlePhase, V^\singlePhase, \NBoldNoSuper^\singlePhase) \nonumber \\
- \sum_{i=1}^n N_{i}^{\incp} \pdBracket{S}{N_i}{U, V, \NBoldNoSuper}(U^\singlePhase, V^\singlePhase, \NBoldNoSuper^\singlePhase) + S(U^{\incp}, V^{\incp}, \NBoldNoSuper^{\incp}).
\end{align}
Using the standard thermodynamic identities
\begin{align} \label{eq:thermo_ids1}
\pdBracket{S}{U}{V, \NBoldNoSuper} = \frac{1}{T}, \quad
\pdBracket{S}{V}{U, \NBoldNoSuper} = \frac{P}{T}, \quad
\pdBracket{S}{N_i}{U, V, \NBoldNoSuper} = -\frac{\mu_i}{T},
\end{align}
into \Cref{eq:deltaS_1}, we get
\begin{align} \label{eq:deltaS_2}
\Delta S = - \frac{U^{\incp}}{T^{\singlePhase}} - \frac{P^{\singlePhase}}{T^{\singlePhase}} V^{\incp} + \sum_{i=1}^n N_{i}^{\incp} \frac{\mu_i^{\singlePhase}}{T^{\singlePhase}} + S(U^{\incp}, V^{\incp}, \NBoldNoSuper^{\incp}).
\end{align}
Furthermore, since entropy is homogeneous of degree one, Euler's theorem gives the incipient phase entropy as:
\begin{align} \label{eq:euler_homogeneous}
S(U^{\incp}, V^{\incp}, \NBoldNoSuper^{\incp})
&= \frac{U^{\incp}}{T^{\incp}} + \frac{P^{\incp}}{T^{\incp}} V^{\incp} - \sum_{i=1}^{n} \frac{\mu_{i}^{\incp}}{T^{\incp}} N_{i}^{\incp}.
\end{align}
Substituting \Cref{eq:euler_homogeneous} into \Cref{eq:deltaS_2} yields:
\begin{align}
\Delta S =
\left(\frac{1}{T^{\incp}} - \frac{1}{T^{\singlePhase}}\right)U^{\incp} + \left(\frac{P^{\incp}}{T^{\incp}} - \frac{P^{\singlePhase}}{T^{\singlePhase}} \right)V^{\incp} - \sum_{i=1}^n N_{i}^{\incp} \left(\frac{\mu_{i}^{\incp}}{T^{\incp}} - \frac{\mu_i^{\singlePhase}}{T^{\singlePhase}}\right).
\end{align}
If \( \Delta S > 0 \), the single-phase state is unstable, and the mixture will split into two phases. Dividing by \(V^{\incp}\), we obtain the famous \textit{tangent plane distance} (TPD) function

\begin{align} \label{eqn:tpd1}
    D(c_1^{\incp}, \dots, c_n^{\incp}, u^{\incp}) &= \frac{\Delta S}{V^{\incp}}, \nonumber\\
      &= u^{\incp} \left(\frac{1}{T^{\incp}} - \frac{1}{T^{\singlePhase}}\right) + \left(\frac{P^{\incp}}{T^{\incp}}  - \frac{P^{\singlePhase}}{T^{\singlePhase}} \right) - \sum_{i=1}^n c_{i}^{\incp} \left(\frac{\mu_{i}^{\incp}}{T^{\incp}} - \frac{\mu_i^{\singlePhase}}{T^{\singlePhase}}\right),
\end{align}
where \( u^{\incp} := U^{\incp}/V^{\incp}\) is the molar internal energy density and \(c_{i}^{\incp} := N_{i}^{\incp}/V^{\incp}\) is molar concentration of \(i^{\text{th}}\) component in the incipient trial phase. Note that, the independent variables are \(u^{\incp}\) and \(c_1^{\incp}, \dots, c_n^{\incp}\). The intensive properties of the incipient phase temperature \(T^{\incp}\), pressure \(P^{\incp}\), and chemical potentials \(\mu_i^{\incp}\) can be obtained as follows. First, the temperature is determined by inverting the internal energy relation:
\[
u^{\incp} = U(T^{\incp}, 1.0, c_1^{\incp}, \dots, c_n^{\incp}).
\]
Once \(T^{\incp}\) is known, the pressure can be computed as:
\[
P^{\incp} = P(T^{\incp}, 1.0, c_1^{\incp}, \dots, c_n^{\incp}),
\]
and the chemical potentials follow similarly:
\[
\mu_i^{\incp} = \mu_i(T^{\incp}, 1.0, c_1^{\incp}, \dots, c_n^{\incp}), \qquad i = 1, \dots, n.
\]

The TPD function \(D\) can be used for stability testing for a system with a specified UVN. If \(D \le 0\) for all admissible \( \{c_{1}^{\incp}, \dots, c_{n}^{\incp}, u^{\incp} \}\), then the system is stable as single-phase. To find out a state \( \{c_{1}^{\incp}, \dots, c_{n}^{\incp}, u^{\incp} \}\) for which \(D > 0\), one seeks to find the local maxima of the function \(D\). Differentiating \(D\) with respect to its independent variables \(u^{\incp}\) and \(\{c_1^{\incp}, \dots, c_n^{\incp}\}\) and applying the Gibbs–-Duhem relation yields (for details, see \cite{smejkal_smejkal_2021, mikyska_investigation_2012})

\begin{align}
\frac{1}{T^{\incp}} - \frac{1}{T^{\singlePhase}} &= 0, \label{eqn:stability_temp_equal}\\
\frac{1}{T^{\singlePhase}}(\mu_i^{\incp} - \mu_i^{\singlePhase}) &= 0, \qquad i = 1, \dots, n. \label{eqn:chem_pot_stability}
\end{align}
\Cref{eqn:stability_temp_equal} implies that the temperature of the incipient phase is equal to that of the reference phase temperature; \Cref{eqn:tpd1} simplifies to
\begin{align} \label{eqn:tpd2}
    D = \left(\frac{P^{\incp}}{T^{\incp}}  - \frac{P^{\singlePhase}}{T^{\singlePhase}} \right) - \sum_{i=1}^n c_{i}^{\incp} \left(\frac{\mu_{i}^{\incp}}{T^{\incp}} - \frac{\mu_i^{\singlePhase}}{T^{\singlePhase}}\right)
\end{align}

Interestingly, this is identical to the TPD function obtained in VTN-stability analysis~\cite{nichita_isochoric_2009}. Therefore, in principle, one can solve a VTN-stability problem instead of a UVN-stability problem. The system of \(n\) equations \eqref{eqn:chem_pot_stability}, can be solved for the incipient phase concentrations \(c_i^{\incp},\; i = 1, \dots, n\) and if the resulting TPD for this set of concentrations is greater than zero, then the system is thermodynamically unstable and will split into multiple phases. Note that the temperature of the incipient phase, \(T^{\incp}\), is the same as the reference phase temperature \(T^{\singlePhase}\). Furthermore, a good initial guess is critical to obtaining convergence of the stability analysis test, which is discussed next.

\subsection{Initial guess for stability analysis}

We employ three different types of initial guesses for the stability analysis test:

\begin{itemize}
    \item Simplex-based approach, as discussed by~\cite{smejkal_phase_2017, kumar_solving_2026}.
    \item Saturation pressure-based approach, as discussed by~\cite{michelsen_isothermal_1982, mikyska_investigation_2012}.
    \item Gaussian random perturbations added to the hypothetical single-phase concentrations.
\end{itemize}

These three sets of initial conditions provide a diverse and well-distributed coverage of the solution space, which improves the robustness of the stability analysis. By exploring multiple starting points, we increase the likelihood of detecting incipient phase formation and obtaining a reliable estimate for the concentrations \(c_1^{\incp}, \dots, c_n^{\incp}\) and molar internal energy density \(u^{\incp}\) of the trial phase. A practical consideration in implementing the stability analysis is to skip the trivial solutions, i.e., solutions where the computed trial phase molar concentration is identical to the input molar concentration.

\subsection{Initial guess from stability results}
A good initial guess is crucial for achieving convergence in flash calculations. Stability analysis provides the trial phase concentration and internal energy density. To initiate the flash calculations, we also need the volumetric phase split. Following Kumar \etal~\cite{kumar_solving_2026}, the trial phase volume is initially set to half of the total volume and then iteratively halved until a phase split is found that yields a two-phase entropy greater than that of the corresponding single-phase state. The energy density and molar concentration can then be scaled by this trial volume to obtain initial estimates of the trial phase internal energy and mole numbers, and the corresponding bulk-phase quantities are determined by subtracting from the total quantities.

\section{Reformulation of the UVN-flash for fluid dynamics}\label{sec:flash}
\begin{figure}[htbp]
    \centering
\begin{tikzpicture}[node distance=1.5cm, every node/.style={font=\small}]
\tikzstyle{startstop} = [rectangle, rounded corners, minimum width=3cm, minimum height=1cm, text centered, draw=black, fill=blue!10]
\tikzstyle{decision} = [diamond, minimum width=3.5cm, minimum height=1cm, text centered, draw=black, fill=orange!20, aspect=2]
\tikzstyle{process} = [rectangle, minimum width=3.5cm, minimum height=1cm, text centered, draw=black, fill=green!20]
\tikzstyle{arrow} = [thick,->,>=stealth]

\node (start) [startstop] {Inputs: $U_{\text{spec}}, V_{\text{spec}}, \mathbf{N}_{\text{spec}}$};

\node (IsSinglePhase) [decision, below of=start, yshift=-1.5cm] {Is single phase ?};
\node (returnSingleIfUserKnowsForSure) [startstop, right of=IsSinglePhase, xshift=4cm] {Return single-phase results};

\node (xguess) [decision, below of=IsSinglePhase, yshift=-1.5cm] {\shortstack{Initial guess\\ provided?}};
\node (FlashWithUserInitialGuess) [process, left of=xguess, xshift=-4cm] {\shortstack{Perform flash with \\user-provided initial guess}};
\node (flashSuccessWithUserGuess) [decision, below of=FlashWithUserInitialGuess, yshift=-0.5cm] {Success?};
\node (FlashResultWithUserGuess) [startstop, below of=flashSuccessWithUserGuess, yshift=-0.5cm] {\shortstack{Return flash result}};

\node (stability) [process, right of=xguess, xshift=4cm] {Run stability analysis};
\node (IsTwoPhaseAfterStabilityTest) [decision, below of=stability, yshift=-1.cm] {Two Phase?};
\node (InitGuessFromStability) [process, below of=IsTwoPhaseAfterStabilityTest, yshift=-0.5cm] {\shortstack{Generate Initial guess \\from stability results}};

\node (SinglePhaseAfterStabilityTest) [startstop, below of=xguess, yshift=-3.5cm] {\shortstack{Return single-phase results}};
\node (PerformFlashWithStabilityGuess) [process, below of=InitGuessFromStability, yshift=-1.0cm] {Perform two-phase flash};
\node (IsTwoPhaseAfterStability) [decision, below of=PerformFlashWithStabilityGuess, yshift=-0.5cm] {Success?};
\node (FlashResultAfterStabilityTest) [startstop, below of=IsTwoPhaseAfterStability, yshift=-1cm] {\shortstack{Return Two-phase results}};
\node (FlashFailedAfterStability) [startstop, left of=IsTwoPhaseAfterStability, xshift=-3.0cm] {Return failed};
\draw [arrow] (start) -- (IsSinglePhase);

\draw [arrow] (IsSinglePhase) -- node[anchor=south] {Yes} (returnSingleIfUserKnowsForSure);
\draw [arrow] (IsSinglePhase.south) -- (xguess.north)  node[midway, anchor=east] {No};
\draw [arrow] (xguess) -- node[anchor=south] {Yes} (FlashWithUserInitialGuess);

\draw [arrow] (FlashWithUserInitialGuess) -- (flashSuccessWithUserGuess);
\draw [arrow] (flashSuccessWithUserGuess) -- node[anchor=west] {Yes} (FlashResultWithUserGuess);
\draw [arrow] (flashSuccessWithUserGuess) -- node[anchor=south] {No} +(8.5,0) |- (stability);
\draw [arrow] (xguess) -- node[anchor=south] {No} (stability);
\draw [arrow] (stability) -- (IsTwoPhaseAfterStabilityTest);
\draw [arrow] (IsTwoPhaseAfterStabilityTest) -- node[anchor=west] {Yes} (InitGuessFromStability);
\draw [arrow] (InitGuessFromStability) -- (PerformFlashWithStabilityGuess);
\coordinate (turn) at ($(IsTwoPhaseAfterStabilityTest.west) + (-0.25,0)$);
\coordinate (drop) at ($(turn -| SinglePhaseAfterStabilityTest.north)$);
\draw [arrow] (IsTwoPhaseAfterStabilityTest) -- (turn) -- (drop) -- (SinglePhaseAfterStabilityTest.north) node[midway, anchor=east] {No};
\draw [arrow] (PerformFlashWithStabilityGuess) -- (IsTwoPhaseAfterStability);
\draw [arrow] (IsTwoPhaseAfterStability) -- (FlashResultAfterStabilityTest) node[midway, anchor=east] {Yes};

\draw [arrow] (IsTwoPhaseAfterStability.west) -- (FlashFailedAfterStability.east) node[midway, anchor=south] {No};
\end{tikzpicture}
\caption{Flow chart for flash calculations}
    \label{fig:flow_chart}
\end{figure}
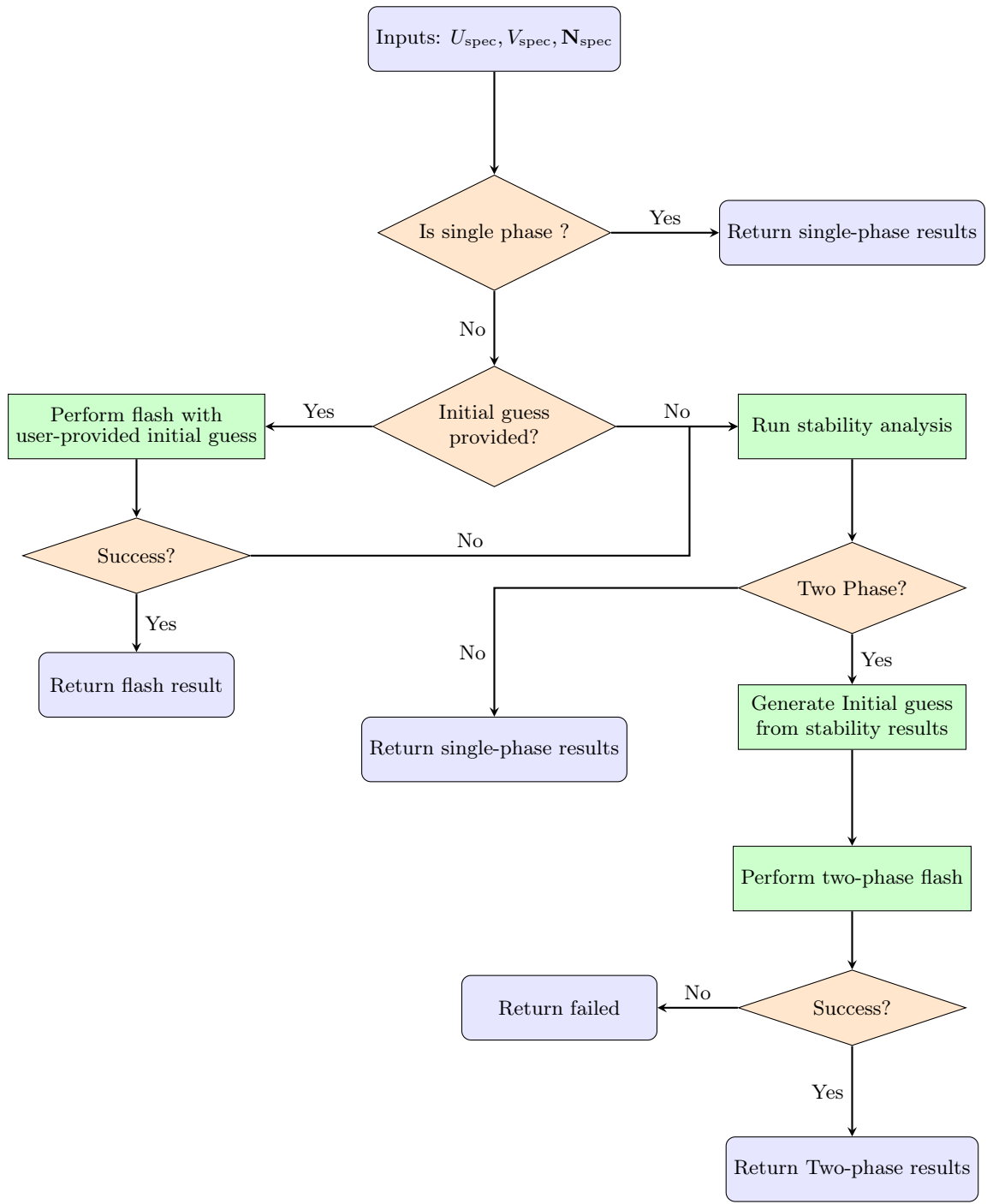

If stability analysis indicates that the mixture will split into two phases, flash is performed to determine the equilibrium phase split. At a given time step in a computational cell, the available fluid variables are the total mass density $\rho$, the specific internal energy $e$, and the overall mixture composition expressed as mole fractions $\moleFrac := \{z_1, \dots, z_n\}$. From these quantities, the total internal energy, \(U^{\total}\) and the total number of moles of \(i^{\scriptstyle\text{th}}\) component, \(N^{\total}_i\) can be computed as
\begin{align}
    U^{\total} &= \rho e V, \\
    N^{\total}_i &= z_i \frac{\rho V}{\sum_j z_j M_{j,w}}, \qquad i = 1,\dots,n,
\end{align}
where $V$ is the cell volume and $M_{j,w}$ denotes the molecular weight of component $j$. The flow diagram for the flash calculation is provided in \Cref{fig:flow_chart}.
In the following subsections, we present the UVN-flash formulation tailored for fluid-dynamical applications.

\subsection{UVN-flash in TVN-Space}
For a closed system whose total internal energy, volume, and mass (i.e., the number of moles) are fixed, the entropy tends to be at a maximum at equilibrium. Let the system exist in two phases, say gas and liquid. The total entropy of the system, $S^{\total}$ can be expressed as

\begin{equation} \label{Flash:entropy_uvn}
    S^{\total} = S(U^{\gas}, V^{\gas}, \NBoldNoSuper^{\gas}) + S(U^{\liq}, V^{\liq}, \NBoldNoSuper^{\liq}).
\end{equation}
The liquid phase quantities can be expressed in terms of the total and the gas phase quantities, i.e. \(U^{\liq} = U^{\total} - U^{\gas}, V^{\liq} = V - V^{\gas}\) and \(N_1^{\liq} = N_1^{\total} - N_1^{\gas}, \dots, N_n^{\liq} = N_n^{\total} - N_n^{\gas}\). \Cref{Flash:entropy_uvn} can be rewritten as
\begin{equation} \label{Flash:entropy_uvn2}
    S^{\total} = S(U^{\gas}, V^{\gas}, \NBoldNoSuper^{\gas}) + S(U^{\total} - U^{\gas}, V - V^{\gas}, \NBoldNoSuper^{\Res}),
\end{equation}
where 
\begin{align} \label{eqn:moles_remaining_phase}
\NBoldNoSuper^{\Res} := \{N_1^{\total} - N_1^{\gas}, \dots, N_n^{\total} - N_n^{\gas}\}.
\end{align} 
The entropy function \eqref{Flash:entropy_uvn2} can be maximized, using the gas--phase variables as optimization variables, to determine the equilibrium phase split. This approach has been employed by Castier~\cite{castier_solution_2009} and Smejkal \etal~\cite{smejkal_phase_2017}. However, the EOS allows us to compute the entropy and other thermodynamic quantities as a function of \(T, V\) and \(\NBoldNoSuper\). Thus, at every sub-iteration of the maximization, the temperature must be obtained by solving 
\begin{align} \label{eqn:internal_energy_tvn}
    U = U(T, V, \NBoldNoSuper), 
\end{align} 
for a given internal energy $U$, volume $V$, and composition $\NBoldNoSuper$. In a pipeline or a compositional reservoir simulation, where thousands or even millions of such computations need to be performed, these additional Newton iterations associated with this inversion step for determining the temperature, can become a bottleneck. This motivates the reformulation of the entropy maximization problem directly in terms of the natural EOS variables \(T, V\) and \(\NBoldNoSuper\). Using the thermodynamic relation \(U = A + TS\), the entropy for a given internal energy \(U^{\total}\) can be expressed as
\begin{align} \label{SQ}
    S^{\scriptscriptstyle\text{TVN}} = \frac{U^{\total} - A^{\scriptscriptstyle\text{TVN}}}{T},
\end{align}
where the superscript $\scriptstyle\text{TVN}$ indicates that the function arguments are \(T, V, N_1, \dots, N_n\) and
\[A^{\scriptscriptstyle\text{TVN}} := \sum_{\substack{k \in \{\gas, \liq\}}} A(T, V^{k}, \NBoldNoSuper^{k}).\]

When the UVN flash is performed in TVN space, the total volume $V$ and mole numbers $N_i^{\total}$ can be poorly scaled: $V$ may range from millimeters to Kilometers and $N_i^{\total}$ can vary several from nearly zero to thousands, depending on the composition.
To improve numerical stability, we instead employ the gas-phase component densities $\boldsymbol{\rho}^{\gas} := \{\rho_1^{\gas}, \dots, \rho_n^{\gas}\}$ together with the gas-phase volumetric fraction $\phaseFrac \in [0,1]$ as primary variables. These variables remain naturally well-scaled across a wide range of compositions and phase splits, thereby improving the robustness of the flash calculations.  
\newline
The gas-phase volume is related to the total volume by
\begin{align}
    V^{\gas} = \phaseFrac V,
\end{align}
and the mole numbers in each phase are obtained from the species densities according to
\begin{align}
    N_i^{\gas} &= \frac{\rho_i^{\gas} V^{\gas}}{M_{i,w}}, \qquad
    N_i^{\liq} = N_i - N_i^{\gas}, \qquad i=1,\dots,n,
\end{align}
with the inverse relation  
\begin{align} \label{rho_i_from_N_V}
\rho_i^{\gas} = \frac{M_{i,w} N_i^{\gas}}{V^{\gas}}.
\end{align}
With these variables, the entropy function can be expressed as
\begin{align} \label{SQRhoAlpha}
    \Q = \frac{U^{\total} - A^{\TAlphaRho}}{T},
\end{align}
where the superscript $\scriptstyle{T\alpha\rho}$ indicates that the function arguments are \(T, \phaseFrac, \rho_1^{\gas}, \dots, \rho_n^{\gas}\) and 
\[A^{\TAlphaRho} := A^{\gas}+A^{\liq}.\]
The phasic Helmholtz energies are defined by
\begin{align}
    A^{\gas} := A(T, \phaseFrac, \rho_1^{\gas}, \dots, \rho_n^{\gas}), \quad A^{\liq} := A(T, \liqFrac, \rho_1^{\liq}, \dots, \rho_n^{\liq}),
\end{align}
where \(\rho_i^{\liq} = (N_i - N_i^{\gas}) M_{i, w}/(V - V^{\gas})\) and \(\liqFrac = 1 - \phaseFrac\). The liquid-phase densities \(\rho_i^{\liq}\) can be expressed in terms of gas-phase variables as

\begin{align} \label{rhoL_rhoG}
\rho_i^{\liq}(\rho_i^{\gas}, \phaseFrac) = \frac{M_{i,w} N_i}{(1-\phaseFrac) V} - \frac{\phaseFrac}{1-\phaseFrac} \, \rho_i^{\gas}, \quad i=1,\dots,n.
\end{align}
Thus, \(A^{\liq}\) can be expressed entirely as a function of \((T, \phaseFrac, \boldsymbol{\rho}^{\gas})\):
\[
A^{\liq} = A\Bigl(T, \alpha^{\liq}(\phaseFrac), \rho_1^{\liq}(\rho_1^{\gas}, \phaseFrac), \dots, \rho_n^{\liq}(\rho_n^{\gas}, \phaseFrac)\Bigr).
\]

From thermodynamics, we know that the entropy should be maximized under the prescribed \(UV\NBoldNoSuper\) constraints, irrespective of the particular choice of variables used to represent the entropy function. Consequently, the critical points of the entropy defined in \Cref{SQRhoAlpha} must coincide with the thermodynamic equilibrium for given  \(UV\NBoldNoSuper\) constraints, as will be verified in the next subsection. The critical points of the entropy function are obtained as: 
\begin{align} \label{eqn:QFunction_stationarity_points}
\pd{\Q}{T} = 0, \qquad
\pd{\Q}{\alpha^{\gas}} = 0, \qquad
\pd{\Q}{\rho_i^{\gas}} = 0 \quad (i=1,\dots,n),
\end{align}
which yields system of \(n+2\) non-linear equations.

\subsection{Thermodynamic consistency check} \label{sec:thermodynamic_consistency}

In this section, we show that the critical points of the entropy function as defined by \Cref{eqn:QFunction_stationarity_points}, coincide with the thermodynamic conditions for phase equilibrium. Specifically, it suffices to show that these critical points correspond to equality of pressure, equality of chemical potential for each component across the phases, a common temperature, and recovery of the prescribed internal energy constraint. The condition of equal temperature is trivially satisfied, as a common temperature is imposed in the formulation from the outset. We recall that
\begin{align} \label{eqn:prelim_partials}
\rho_i^{\gas} = \frac{N_i^{\gas} M_{i, w}}{V^{\gas}} = \frac{N_i^{\gas} M_{i, w}}{\alpha^{\gas} V} \implies \pd{\rho_i^{\gas}}{N_i^{\gas}} = \frac{M_{i, w}}{V^{\gas}}, \quad \pd{\rho_i^{\gas}}{\alpha^{\gas}} = -\frac{1}{\alpha^{\gas}} \frac{N_i^{\gas} M_{i, w}}{\alpha^{\gas} V} = -\frac{\rho_i^{\gas}}{\alpha^{\gas}} .
\end{align}
Furthermore, the inverse relations are given by
\begin{align}  \label{eqn:rho_partials}
    \pd{N_i^{\gas}}{\rho_i^{\gas}} = \frac{V^{\gas}}{M_{i, w}}, \quad \pd{\alpha^{\gas}}{\rho_i^{\gas}} = -\frac{\alpha^{\gas}}{\rho_i^{\gas}}, \quad \pd{V^{\gas}}{\rho_i^{\gas}} = -\frac{V^{\gas}}{\rho_i^{\gas}}.
\end{align}
1) The stationarity condition with respect to the mass densities reads:
\begin{align}
    0
    &= \pdBracket{\Q(T, \phaseFrac, \rho_1^{\gas},\dots,\rho_n^{\gas})}{\rho_i^{\gas}}
               {T,\phaseFrac,\{\rho_j^{\gas}\}_{j\ne i}}
        \qquad\nonumber\\
    &= -\frac{1}{T}\,
       \pdBracket{A^{\TAlphaRho}(T,\phaseFrac,\rho_1^{\gas},\dots,\rho_n^{\liq})}
               {\rho_i^{\gas}}{T,\phaseFrac,\{\rho_j^{\gas}\}_{j\ne i}} \label{eqn:partial_Q_partial_rho_i}.
\end{align}
We therefore focus on the derivative of the total Helmholtz free energy with respect to \(\rho_i^{\gas}\).
Using \(A^{\TAlphaRho}=A^{\gas}+A^{\liq}\), we obtain
\begin{align} \label{eqn:partial_A_tot_rho_gas_tmp}
    \pdBracket{A^{\TAlphaRho}}{\rho_i^{\gas}}{T,\phaseFrac,\{\rho_j^{\gas}\}_{j\ne i}}
    &= \pdBracket{A^{\gas}}{\rho_i^{\gas}}{T,\phaseFrac,\{\rho_j^{\gas}\}_{j\ne i}}
      + \pdBracket{A^{\liq}}{\rho_i^{\gas}}{T,\phaseFrac,\{\rho_j^{\gas}\}_{j\ne i}}.
\end{align}
As $\rho_i^{\gas}$ is a function of $N_i^{\gas}$ and \(V^{\gas}\) (see \Cref{eqn:prelim_partials}); using the chain rule gives
\begin{align}
    \pdBracket{A^{\gas}}{\rho_i^{\gas}}{}
    &= \pd{V^{\gas}}{\rho_i^{\gas}} \pdBracket{A^{\gas}}{V^{\gas}}{T,\{N_k^{\gas}\}}  + \sum_k \pd{N_k^{\gas}}{\rho_i^{\gas}} \pdBracket{A^{\gas}}{N_k^{\gas}}{T,V^{\gas},\{N_{m}^{\gas}\}_{m\ne k}}, \label{eqn:summation_A_rho_g} \\
    \pdBracket{A^{\liq}}{\rho_i^{\gas}}{}
    &= \pd{V^{\liq}}{\rho_i^{\gas}} \pdBracket{A^{\liq}}{V^{\liq}}{T,\{N_k^{\liq}\}}  + \sum_k \pd{N_k^{\liq}}{\rho_i^{\gas}} \pdBracket{A^{\liq}}{N_k^{\liq}}{T,V^{\liq},\{N_{m}^{\liq}\}_{m\ne k}}. \label{eqn:summation_A_rho_l}
\end{align}
In equations \eqref{eqn:summation_A_rho_g} and \eqref{eqn:summation_A_rho_l}, only terms inside the summation where \(k=i\) contribute. Using the chemical potential definition,
\(\mu_i^{\gas}=\pdBracket{A^{\gas}}{N_i^{\gas}}{T,V^{\gas},\{N_{j}^{\gas}\}_{j\ne i}}\) and rearranging, we get

\begin{align} \label{eqn:partial_A_rho_gas_tmp}
    \pdBracket{A^{\TAlphaRho}}{\rho_i^{\gas}}{}
    &= \mu_i^{\gas}\,\pd{N_i^{\gas}}{\rho_i^{\gas}}
      + \mu_i^{\liq}\,\pd{N_i^{\liq}}{\rho_i^{\gas}}
      + \pd{V^{\gas}}{\rho_i^{\gas}}\underbrace{\pdBracket{A^{\gas}}{V^{\gas}}{T, \{N_k^{\gas}\}}}_{-p^{\gas}} 
      + \pd{V^{\liq}}{\rho_i^{\gas}}\underbrace{\pdBracket{A^{\liq}}{V^{\liq}}{T, \{N_k^{\liq}\}}}_{-p^{\liq}}.
\end{align}
Since \(V^{\liq}=V - V^{\gas}\), it follows that \(\pd{V^{\liq}}{\rho_i^{\gas}} = -\pd{V^{\gas}}{\rho_i^{\gas}}\). From \Cref{eqn:prelim_partials}, we have \[\pd{V^{\gas}}{\rho_i^{\gas}} = -V^{\gas}/\rho_i^{\gas}, \quad \pd{N_i^{\gas}}{\rho_i^{\gas}}=V^{\gas}/M_{i,w}.\] Moreover, using mass balance relation, \(N_i^{\liq}=N_i-N_i^{\gas}\), we obtain
\(\pd{N_i^{\liq}}{\rho_i^{\gas}}=-V^{\gas}/M_{i,w}\). Substituting these expressions in \Cref{eqn:partial_A_rho_gas_tmp} and using \Cref{eqn:partial_A_tot_rho_gas_tmp} gives
\begin{align}
    \pdBracket{A^{\TAlphaRho}}{\rho_i^{\gas}}{}
    &= \mu_i^{\gas}\frac{V^{\gas}}{M_{i,w}} - \mu_i^{\liq}\frac{V^{\gas}}{M_{i,w}}
      + \left(p^{\gas} - p^{\liq}\right) \frac{V^{\gas}}{\rho_i^{\gas}}, \nonumber\\
    &= \left(\mu_i^{\gas}-\mu_i^{\liq}\right)\frac{V^{\gas}}{M_{i,w}} + \left(p^{\gas} - p^{\liq}\right) \frac{V^{\gas}}{\rho_i^{\gas}}.
\end{align}
The stationarity condition \eqref{eqn:partial_Q_partial_rho_i} then yields:
\begin{align} \label{eqn:partial_Q_partial_rho_i_2}
    0 = (\mu_i^{\gas}-\mu_i^{\liq})\frac{\rho_i^{\gas}}{M_{i,w}} + (p^{\gas} - p^{\liq}), \qquad \forall i \in \{1, \dots, n\}.
\end{align}
Note that \Cref{eqn:partial_Q_partial_rho_i_2} is system of \(n\) equations, which in expanded form reads:
\begin{subequations} \label{eqn:partial_A_partial_density_final}
    \begin{align}
        (\mu_1^{\gas}-\mu_1^{\liq})\frac{\rho_1^{\gas}}{M_{1,w}} &+ (p^{\gas} - p^{\liq}) = 0, \\
        &\vdots \\
        (\mu_n^{\gas}-\mu_n^{\liq})\frac{\rho_n^{\gas}}{M_{n,w}} &+ (p^{\gas} - p^{\liq}) = 0. 
    \end{align}
\end{subequations}
2) The stationarity condition with respect to the gas phase fraction \(\phaseFrac\) reads:
\begin{align}
\pdBracket{A^{\TAlphaRho}(T, \phaseFrac, \rho_1^{\gas}, \dots, \rho_n^{\gas})}{\phaseFrac}{T, \rho_1^{\gas}, \dots, \rho_n^{\gas}} &= 0.
\end{align}
Substitute \(A^{\TAlphaRho} = A^{\gas} + A^{\liq}\) and differentiate with respect to \(\phaseFrac\), we get
\begin{align}
\pd{A^{\TAlphaRho}}{\phaseFrac} &= \pd{A^{\gas}}{\phaseFrac} + \pd{A^{\liq}}{\phaseFrac}.
\end{align}
Since \(\phaseFrac = V^{\gas} / V\), the partial derivative with respect to \(\phaseFrac\) is
\begin{align}
\pd{A^{\gas}}{\phaseFrac} 
=  \pd{A^{\gas}}{V^{\gas}} \pd{V^{\gas}}{\phaseFrac} = \frac{1}{V} \pd{A^{\gas}}{V^{\gas}}.
\end{align}
For the liquid phase, using \(V^{\liq} = V - V^{\gas}\), we obtain
\begin{align}
\pd{A^{\liq}}{\phaseFrac} 
&= \pd{A^{\liq}}{V^{\liq}} \pd{V^{\liq}}{\phaseFrac} = -\frac{1}{V} \pd{A^{\liq}}{V^{\liq}} .
\end{align}
Combining the two and recalling \(\pd{A}{V} = -p\), it follows that
\begin{align}
\pd{A^{\TAlphaRho}}{\phaseFrac} 
&= \frac{p^{\liq} - p^{\gas}}{V}.
\end{align}
At the stationarity point \(\pd{A^{\TAlphaRho}}{\phaseFrac}  = 0\), which implies
\begin{align} \label{eqn:mech_eqm}
\boxed{
 p^{\gas} = p^{\liq}}, 
\end{align}
corresponding to the condition of mechanical equilibrium.
\newline\\
Substituting \(p^{\gas} = p^{\liq}\) in \Cref{eqn:partial_A_partial_density_final} yields
\begin{align} \label{eqn:chem_eqm}
\boxed{
    \mu_i^{\gas} = \mu_i^{\liq}, \quad \forall \in \{1, \dots, n\}},
\end{align}
which is the condition of chemical equilibrium.
\newline\\
3) Stationarity with respect to temperature \(T\) reads:
\begin{align}
\pdBracket{S^{\TAlphaRho}(T, \phaseFrac, \rho_1^{\gas}, \dots, \rho_n^{\gas})}{T}{\phaseFrac, \rho_1^{\gas}, \dots, \rho_n^{\gas}} &= 0.
\end{align}
Recall that \(\Q = (U^{\total} - A^{\TAlphaRho})/T\). Differentiating with respect to \(T\) gives:
\begin{align} \label{eqn:partial_S_partial_T_tmp}
\pd{\Q}{T} &= -\frac{U^\total}{T^2} - \frac{1}{T} \pd{A^{\TAlphaRho}}{T} + \frac{A^{\TAlphaRho}}{T^2}.
\end{align}
Substituting \(A^{\TAlphaRho} = A^{\gas} + A^{\liq}\)  and differentiating with respect to \(T\) yields:
\begin{align}
\pd{A^{\TAlphaRho}}{T} 
&= \pd{A^{\gas}}{T} + \pd{A^{\liq}}{T}.
\end{align}
Using thermodynamic identity
\(
\pd{A}{T} = -S,
\)
we obtain
\begin{align}
\pd{A^{\TAlphaRho}}{T} 
&= -S^{\gas} - S^{\liq}.
\end{align}
Substituting this in \Cref{eqn:partial_S_partial_T_tmp}, we obtain
\begin{align}
\pd{\Q}{T} &= -\frac{U^\total}{T^2} + \frac{S^{\gas} + S^{\liq}}{T} + \frac{A^{\gas} + A^{\liq}}{T^2}, \\
&= \frac{A^{\gas} + TS^{\gas} + A^{\liq} + TS^{\liq} - U^\total}{T^2}.
\end{align}
Applying the relation \(U = A + TS\) for each phase yields
\begin{align} 
\pd{\Q}{T} = \frac{U^{\gas} + U^{\liq} - U^\total}{T^2}. 
\end{align}
The stationary condition \(\pd{\Q}{T} = 0\) therefore gives
\begin{align} \label{eqn:constraint_recovery}
\boxed{
U^\total = U^{\gas} + U^{\liq}
},
\end{align}
which is the desired constraint on the internal energy. Equations \eqref{eqn:mech_eqm}, \eqref{eqn:chem_eqm}, and \eqref{eqn:constraint_recovery} together constitute the conditions of thermodynamic equilibrium under UVN specifications.
\section{Temporal Discretization} \label{sec:time_discretization}

We consider the differential-algebraic system (DAE) that arises from the pipe flow equations \eqref{eq:pipe:vector_form} and tank model \eqref{eq:tank:vector_form} along with the flash problem (\eqref{eqn:QFunction_stationarity_points}). With the state variables partitioned into differential (conservative) variables \( \discreteU \) and algebraic (non-conservative) variables \( \discreteV \), the resulting DAE system can be expressed as
\begin{align}
    \frac{\rd \discreteU}{\rd t} &= \discreteF(\discreteU,\discreteV), \label{dae_eqns_f} \\
    0 &= g(\discreteU, \discreteV) \label{dae_eqns_g}.
\end{align}
The precise form of \(\discreteU\) and \(\discreteV\) for the pipe and tank model is discussed below.

\subsubsection*{\textbf{Pipe flow}} 
For spatially discretized pipe flow, we define:
\begin{align*}
&\discreteU = [\discreteU_{1}, \ldots, \discreteU_{N}]^T, \qquad \discreteU_{i} = [\rho_i, (\rho u)_i, (\rho E)_i]^T,    \\
&\discreteV = [\discreteV_{1}, \ldots, \discreteV_{N}]^T, \qquad \discreteV_{i} = [\rho^{\gas}_{k,i}, \alpha^{\gas}_{i}, T_{i}]^T,
\end{align*}
where \(\rho^{\gas}_{k} \), \( \alpha^{\gas} \), and \( T \) are the partial mass density of component \(k\) in gas phase, volumetric phase fraction of gas, and temperature, respectively. The subscript \(i\) refers to the the \(i^{\text{th}}\) cell. A first-order spatial discretization of the conservation laws yields the semi-discrete form:
\[
\discreteF_i(\discreteU, \discreteV) := -\frac{1}{\Delta x} \left( \Hat{\EulerFlux}_{i+\frac{1}{2}}(\discreteU_{i+1}, \discreteU_i, \discreteV_{i+1}, \discreteV_i) - \Hat{\EulerFlux}_{i-\frac{1}{2}}(\discreteU_i, \discreteU_{i-1}, \discreteV_i, \discreteV_{i-1}) \right),
\]
where \( \Hat{\EulerFlux}_{i \pm \frac{1}{2}} \) denotes the numerical flux function, and \( \discreteF_i \) represents the discretized right-hand side of the system \eqref{Eqns:NS}.

\subsubsection*{\textbf{Tank model}}
For the tank, the differential variables are defined as:
\[
\discreteU = [N_1, \dots, N_n, U]^T,
\]
where \( N_i \) denotes total number of moles of component \(i\) in the mixture and \( U \) is the total internal energy. The algebraic variables \( \discreteV \) are defined as: 
\[
\discreteV = [N_1^{\gas}, \dots, N_n^{\gas}, V^{\gas}, T]^T,
\]
where \( N_i^{\gas}\) is the number of moles of component \( i \) in the gas phase, \( V^{\gas}\) is the volume occupied by the gas phase, and \( T \) is the temperature.
The function \( \discreteF \) for the tank is defined by \Cref{eq:tank_rhs}. 
\newline\newline
\textbf{Time integration scheme}

To integrate the DAE system \eqref{dae_eqns_f} in time, we employ a half-explicit forward Euler scheme \cite{hairer_numerical_1989}. This method consists of an explicit update for the differential variables, followed by a nonlinear solve to enforce the algebraic constraints. First, the differential update is computed as:
\[
\discreteU^{n+1} = \discreteU^n + \Delta t \, \discreteF(\discreteU^n, \discreteV^n),
\]
followed by the solution of the nonlinear constraint:
\[
0 = g(\discreteU^{n+1}, \discreteV^{n+1}),
\]
to determine \( \discreteV^{n+1} \). Solving this constraint corresponds to a flash calculation as formulated by system of equations \eqref{eqn:QFunction_stationarity_points}, in \Cref{sec:flash}.

\section{Numerical Experiments} \label{sec:results}

In this section, we discuss the results. First, we discuss the results for the tank model, followed by the results for the pipeline depressurization. We use PR-EOS for all the results; see \ref{app:PR_EOS} for details. The diagonal elements of the binary interaction parameters (BIP) matrix of PR are set to zero, and the off-diagonal terms are specified in the tables \ref{tab:castier_dynamic_problems} and \ref{tab.pipe.shocktube.composition}.

\begin{table}[htbp]
\centering
\caption{Initial conditions, stream data for Castier Problems 1 and 2}
\label{tab:castier_dynamic_problems}
\begin{tabular}{lcccc}
\hline
\textbf{Quantity} & \textbf{Units} & \textbf{Problem 1} & \textbf{Problem 2} \\
\hline
Components(in order) & -- & \{CH$_4$, H$_2$S\} & \{CO$_2$, C$_{12}$H$_{26}$, C$_{13}$H$_{28}$, C$_{14}$H$_{30}$, C$_{15}$H$_{32}$\} \\
\hline
\multicolumn{4}{l}{\textit{Tank initial conditions}} \\
\hline
Pressure $P_0$        & MPa     & $0.10106$ & $0.001$ \\
Temperature $T_0$     & K      & $300.0$              & $373.15$ \\
Total moles $\mathbf{N}_0$ & mol    & $\{500,\;500\}$         & $\{1\times10^{-8},\;0.1,\;0.6,\;0.2,\;0.1\}$ \\
Volume $V$            & m$^3$  & $24.5708$             & $4.714\times10^{-4}$ \\
Initial phase state   & --     & Single-phase          & Two-phase \\
\hline\\
\multicolumn{4}{l}{\textit{Inlet stream}} \\
\hline
Pressure $P_{\mathrm{in}}$ & MPa     & $5$    & $20$ \\
Temperature $T_{\mathrm{in}}$ & K   & $300.0$              & $310.0$ \\
Molar flow $\mathbf{f}_{\mathrm{in}}$ & mol & $\{4.0,\;6.0\}$ & $\{1\times10^{-2},\; 0.0,\;0.0,\;0.0,\;0.0\}$ \\
Heat input $\dot{Q}$ & J/s   & $0.0$                & $-100.0$ \\
Phase state (inlet)     & --     & Two-phase             & Single-phase \\
\hline
\textit{Outlet stream(s)} & & None & None \\
\hline\\
\textit{BIP} ($\{\delta_{ij}\}_{i \ne j}$) & & 0.083 & 0.0 \\
\hline
\end{tabular}
\end{table}

\subsection{Tank model}
We validate our approach on the two problems from Castier~\cite{castier_dynamic_2010}. The specifications of the problem are summarized in \Cref{tab:castier_dynamic_problems}
\subsubsection{Problem 1: Light components}
\begin{figure}[htbp]
    \centering
    \begin{subfigure}[b]{0.48\textwidth}
        \includegraphics[width=\textwidth]{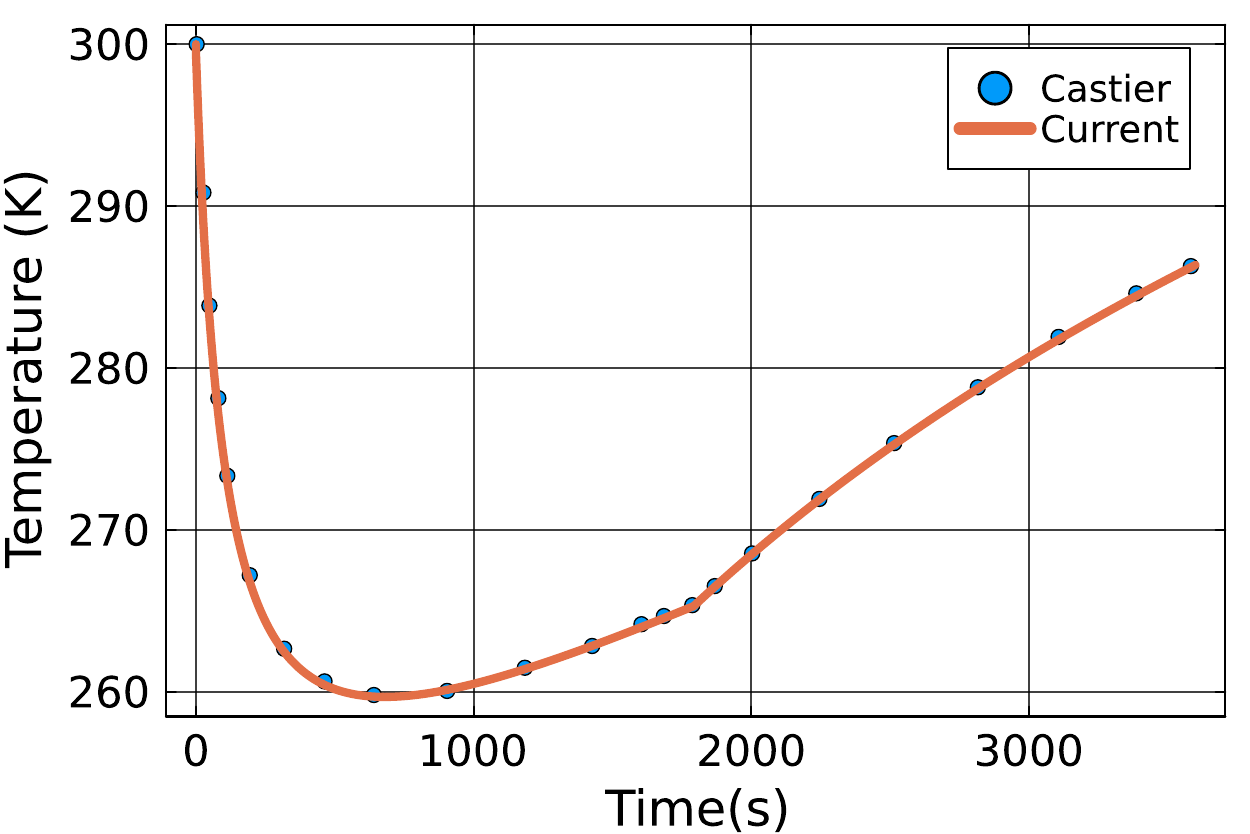}
        \caption{Temporal temperature variation }
        \label{fig.tank.castier.prob1.Tt}
    \end{subfigure}
    \hfill
    \begin{subfigure}[b]{0.48\textwidth}
        \includegraphics[width=\textwidth]{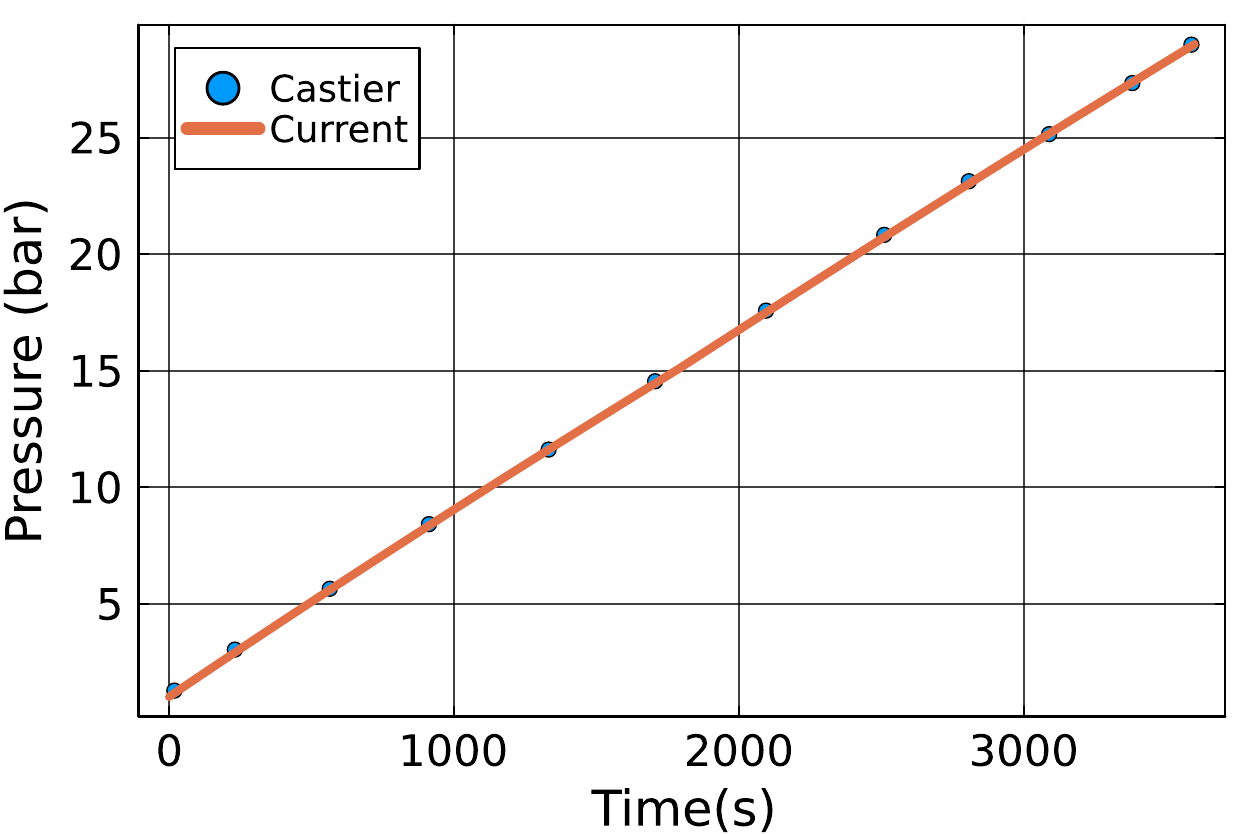}
        \caption{Temporal pressure variation}
        \label{fig.tank.castier.prob1.Pt}
    \end{subfigure}
    \caption{Tank simulation results for Castier~\cite{castier_dynamic_2010} Problem 2. Final $t = 3600s$}
    \label{fig.tank.castier.prob1.results}
\end{figure}

In this problem, we consider a mixture of methane(\ce{CH4}) and hydrogen sulfide (\ce{H2S}).  The fluid in the tank is in a single-phase state at \(t=0s\). There is one input stream feeding the tank, which is in a two-phase state. Since the stream conditions are given in terms of temperature and pressure, we need to perform a PTN-flash~\cite{michelsen_isothermal_1981} only once at the beginning of the simulation to determine the phase split and hence, the enthalpy of individual phases. Total enthalpy is obtained by adding up these individual phasic contributions. The results are shown in \Cref{fig.tank.castier.prob1.results}. An excellent match with the results reported in the Castier~\cite{castier_dynamic_2010} is obtained. Initially, there is a sharp drop in the temperature reaching a minimum temperature (\(\approx 260K\)) around \(680s\) after which the temperature starts to rise and there is an onset of the second phase around \(1792s\). We see a change in the temperature slope at this point. Pressure, on the other hand, keeps on increasing linearly as more mass is being added to the tank.

\subsubsection{Problem 2: \ce{CO2} loading}
\begin{figure}[htbp]
    \centering
    \begin{subfigure}[b]{0.48\textwidth}
        \includegraphics[width=\textwidth]{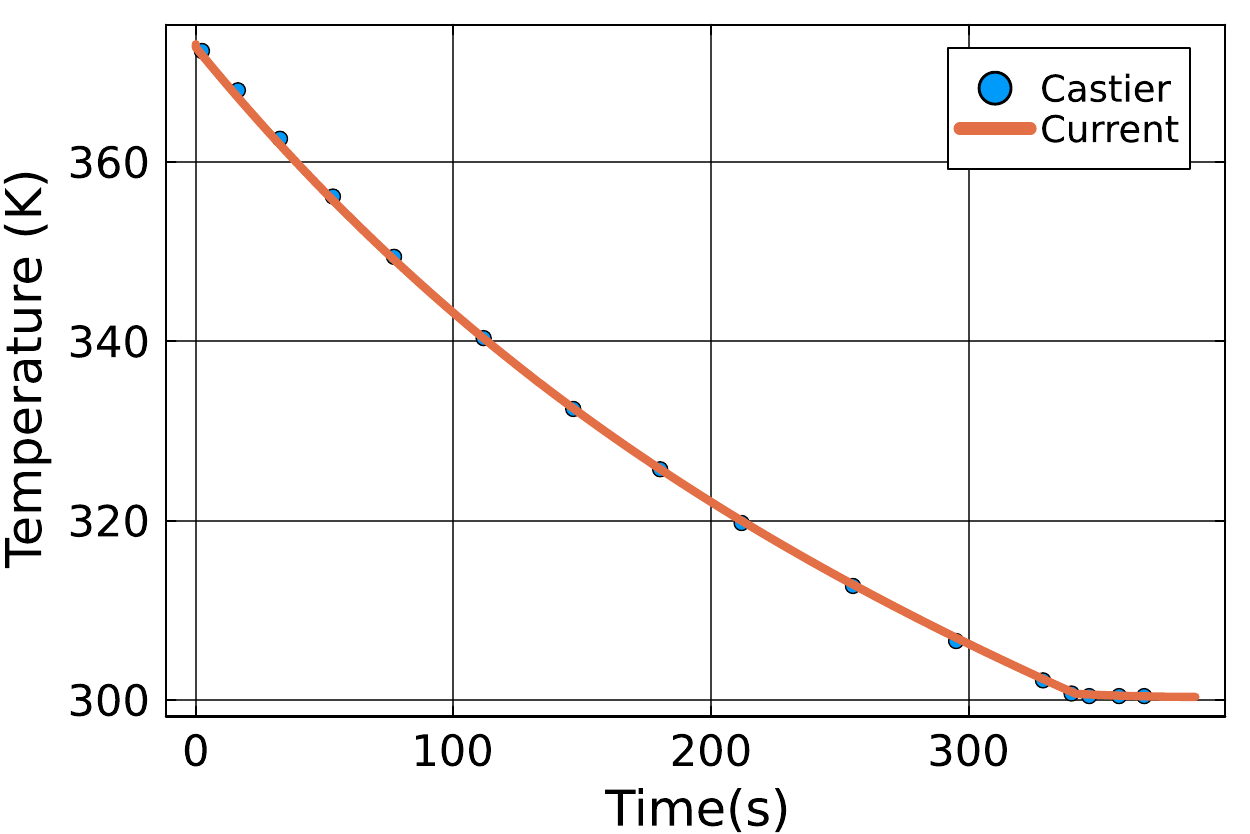}
        \caption{Temporal temperature variation }
        \label{fig.tank.castier.prob2.Tt}
    \end{subfigure}
    \hfill
    \begin{subfigure}[b]{0.48\textwidth}
        \includegraphics[width=\textwidth]{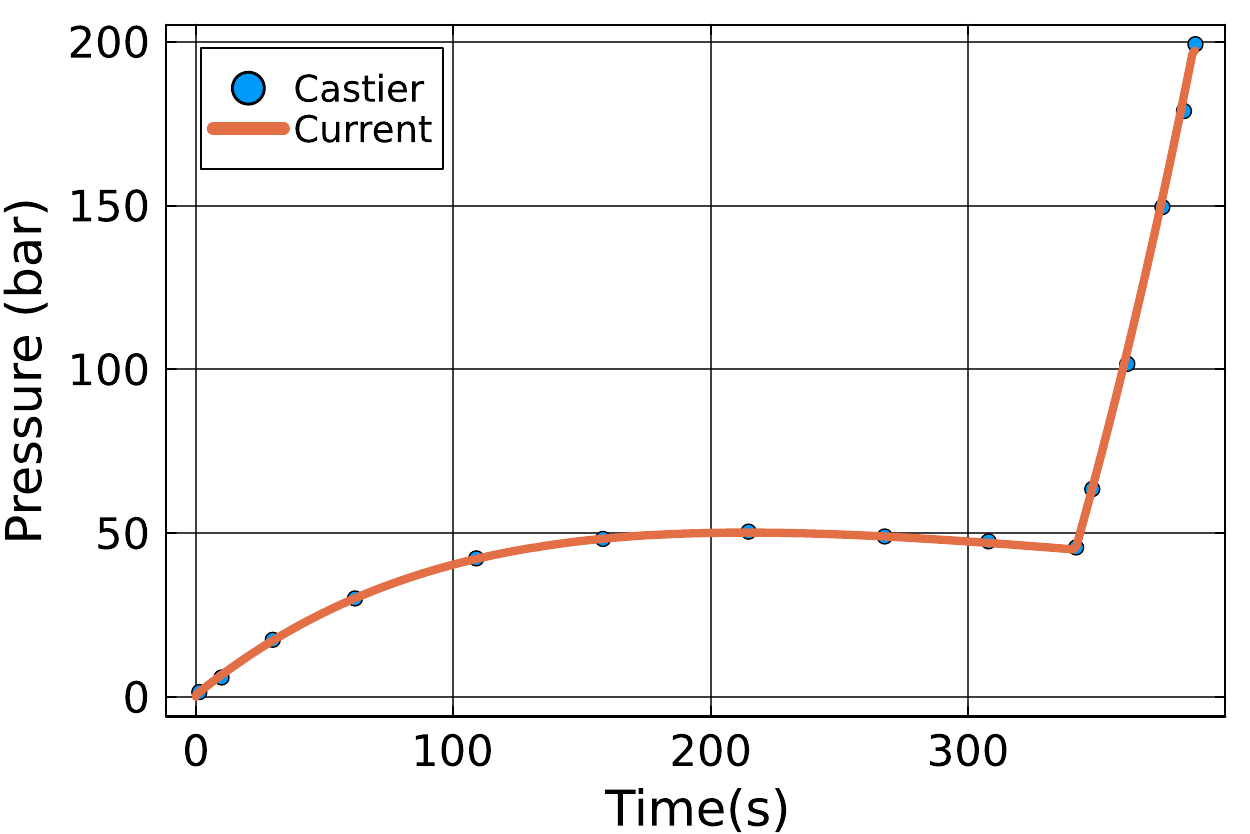}
        \caption{Temporal pressure variation}
        \label{fig.tank.castier.prob2.Pt}
    \end{subfigure}
    \caption{Tank simulation results for Castier~\cite{castier_dynamic_2010} Problem 2. Final $t = 388.3s$}
    \label{fig.tank.castier.prob2.results}
\end{figure}

In this example, we have a mixture of medium-heavy hydrocarbons. Initially, there is a very small amount of \ce{CO2} in the tank. The tank is being fed with \ce{CO2}. For a given pressure, temperature, and feed composition, we can solve the PR--EOS for volume and thereby obtain the enthalpy flow rate. Initially, the fluid in tank is in two-phase state. Throughout the process, the heat is removed at a constant rate of 100 J/s. \Cref{fig.tank.castier.prob2.results}  shows the results obtained in the current work and compares them with those by Casiter~\cite{castier_dynamic_2010}. Again, a very good match can be observed. Until around \(341s\), there are two phases in the tank; beyond this point, only a single phase remains. The temperature then becomes relatively constant, but the pressure exhibits a sharp increase, indicating that the fluid has transitioned to a liquid state, which behaves almost as an incompressible fluid.

To summarize, for both tank depressurization problems, our approach was able to handle the transition from single-phase to two-phase and vice versa without any difficulty. We now turn to a more challenging scenario and apply the approach to pipeline depressurization.

\subsection{Pipeline Depressurization}
Having validated our flash solver for transient simulation for the tank model, we now apply it to pipeline transport of a two-phase multicomponent fluid. We consider six different mixtures: 4 two-component mixtures; a five-component mixture; and a single-component fluid. Such numerical experiments have been previously considered in previous studies; see references~\cite{hammer_method_2013, giljarhus_solution_2012, kumar_new_2025} for single-component fluids and Munkejord \etal~\cite{munkejord_depressurization_2015} for multicomponent mixtures. 

In Munkejord \etal~\cite{munkejord_depressurization_2015}, an outflow boundary condition is prescribed at the open end(right) of the pipe. In contrast, in our simulations, we avoid imposing this boundary condition directly. Instead, we extend the computational domain to twice the original pipe length and formulate the problem as a shock tube with the initial discontinuity located at the midpoint. The reason for adopting this approach is that the flow conditions at the open end will be choked, and hence, the information from outside the pipe cannot propagate upstream into the computational domain. The simulation is terminated before the fastest traveling wave reaches the end of the pipe.

In all simulations, a grid of $800$ cells was used (i.e., $400$ cells per side), whereas Munkejord \etal~\cite{munkejord_depressurization_2015} used $2400$ cells. Furthermore, we employed a CFL value of \(0.9\), while Munkejord \etal~\cite{munkejord_depressurization_2015} used a CFL of $0.85$. Due to the presence of initial discontinuities, an initial time-step of \(1 \times 10^{-12}\) is used to begin the simulation. Subsequent time steps are determined based on the fastest traveling waves:

\begin{align} \label{simulation.dt}
    \Delta t = \frac{\Delta x}{\max_{i} (|u_i \pm a_i|)},
\end{align}
where subscript \(i\) corresponds to the \(i^{\mathrm{th}}\) cell and \(a_i\) represents the speed of sound. The speed of sound is computed using Wood's formula; further details are provided in \ref{sec:woods_pr}.

The initial conditions are provided in \Cref{tab.pipe.shocktube.ic} and the mixture composition is given in \Cref{tab.pipe.shocktube.composition}. To obtain the density and the number of moles, the compressibility factor \(Z\) is first computed. The total number of moles \(N\) is then calculated using the relation \[Z = \frac{PV}{NRT},\] where \(V\) is the total volume of the computational cell, \(R\) is the universal gas constant. The mole vector is subsequently obtained by multiplying \(N\) by the composition vector specified in the \Cref{tab.pipe.shocktube.composition}.

\begin{table}[h!]
\centering
\caption{Summary of Riemann problem initial conditions}
\label{tab.pipe.shocktube.ic}
\begin{tabular}{c l l l l l l}
\hline
\textbf{Case} & \textbf{Fluids} & \textbf{$L$ [m]} & 
\multicolumn{2}{c}{\textbf{Pressure [MPa]}} & \multicolumn{2}{c}{\textbf{Temperature [K]}} \\
\cline{4-5} \cline{6-7}
 &  &  & \textbf{Left} & \textbf{Right} & \textbf{Left} & \textbf{Right} \\
\hline
\vspace{0.3cm}
1 & \{\coo, \methane\}   & 200.0   & 28.568 & 2.0 & 313.65 & 293.15 \\
\vspace{0.0cm}
2 & \{\coo, \hydrogen, \nitrogen, & 288.0   & 12.051 & 2.5 & 283.15 & 283.15 \\
\vspace{0.3cm}
& \oxygen, \methane\} \\
\vspace{0.3cm}
3 & \{\coo\}        & 200.0   & 10.0  & 3.0    & 300.0 & 300.0 \\
4 & \{\coo, \nitrogen\}   & 283.8   & 11.99  & 2.0 & 292.65 & 291.55 \\
5 & \{\coo, \nitrogen\}   & 283.8   & 12.08  & 2.0 & 292.85 & 292.65 \\
6 & \{\coo, \nitrogen\}   & 283.8   & 12.0  & 2.0 & 290.45 & 292.35 \\
\hline
\end{tabular}
\end{table}

\begin{table}[h!]
\centering
\caption{Compositions for different problems}
\label{tab.pipe.shocktube.composition}
\begin{tabular}{c l l c c c c c c}
\hline
\textbf{Case} & \textbf{$\alpha_L$} & \textbf{$\alpha_R$} & \textbf{BIP ($\{\delta_{ij}\}_{i \ne j}$)} & \textbf{\coo} & \methane & \hydrogen & \nitrogen & \oxygen \\
\hline
1 & 0.0 & 1.0 & 0.15 & 0.726 & 0.264 & - & - & - \\
2 & 0.0 & 1.0 & 0.0 & 0.9103 & 0.0115 & 0.04 & 0.0187 & 0.0195\\
3 & 0.0 & 1.0 & -- & 1.0 & - & - & - & - \\
4 & 0.0 & 1.0 & -0.041 & 0.9 & - & - & 0.1 & -\\
5 & 0.0 & 1.0 & -0.041 & 0.8 & - & - & 0.2 & -\\
6 & 0.0 & 1.0 & -0.041 & 0.7 & - & - & 0.3 & -\\
\hline
\end{tabular}

\end{table}
\subsection{Validation}
\begin{figure}[htbp]
    \centering
    \begin{subfigure}[b]{0.48\textwidth}
        \includegraphics[width=\textwidth]{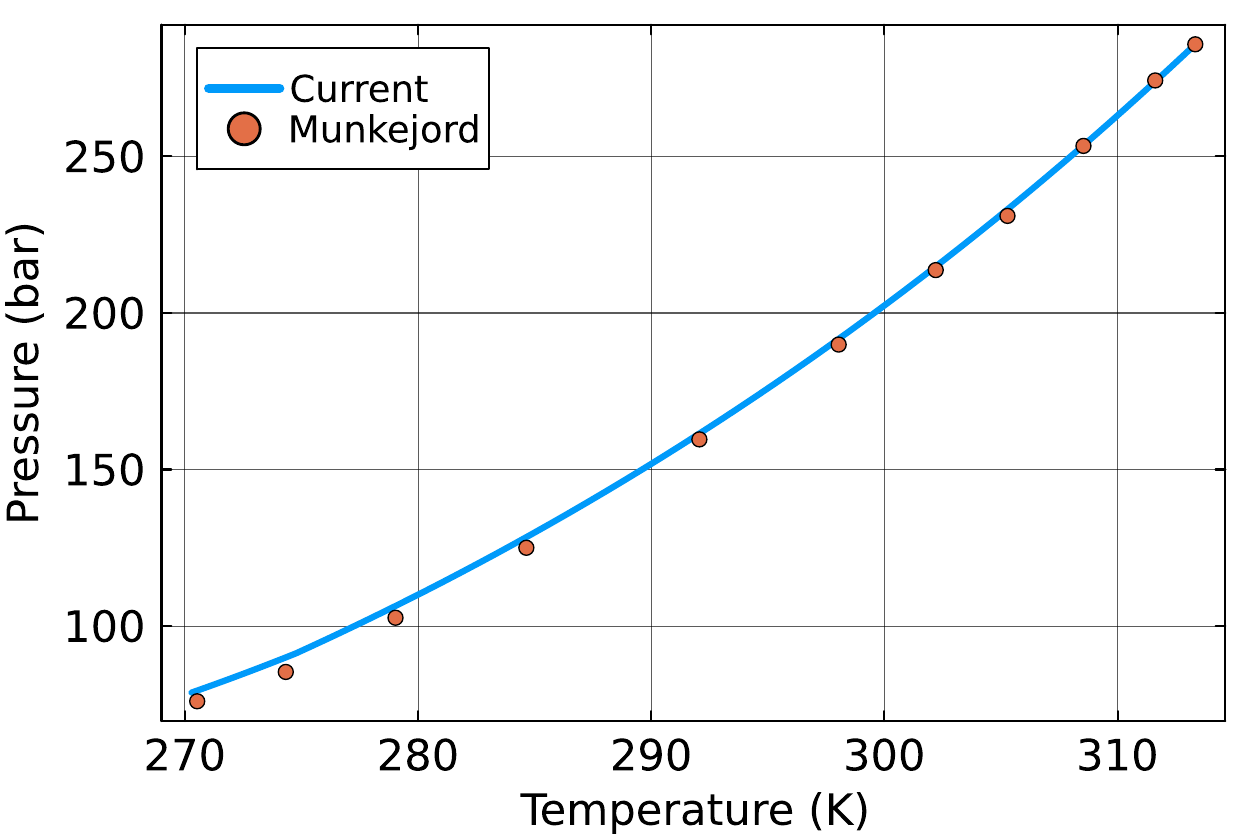}
        \caption{\coo and \methane mixture at \(t = 50\)ms }
        \label{fig.pipe.two_comps.800.HEMTrajectory.PT}
    \end{subfigure}   
    \begin{subfigure}[b]{0.48\textwidth}
        \includegraphics[width=\textwidth]{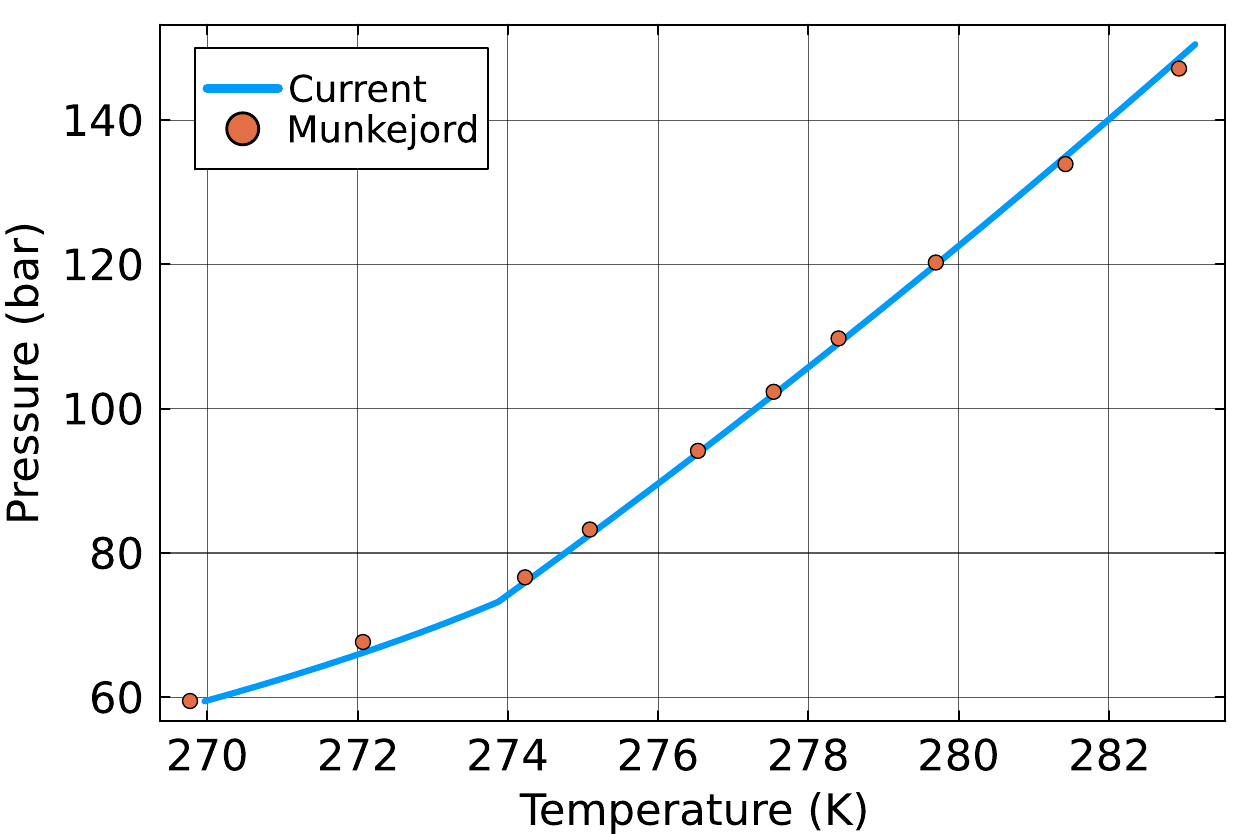}
        \caption{Five component mixture at \(t=50\) ms. }
        \label{fig.pipe.five_comps.800.HEMTrajectory.PT}
    \end{subfigure}  
    \begin{subfigure}[b]{0.48\textwidth}
        \includegraphics[width=\textwidth]{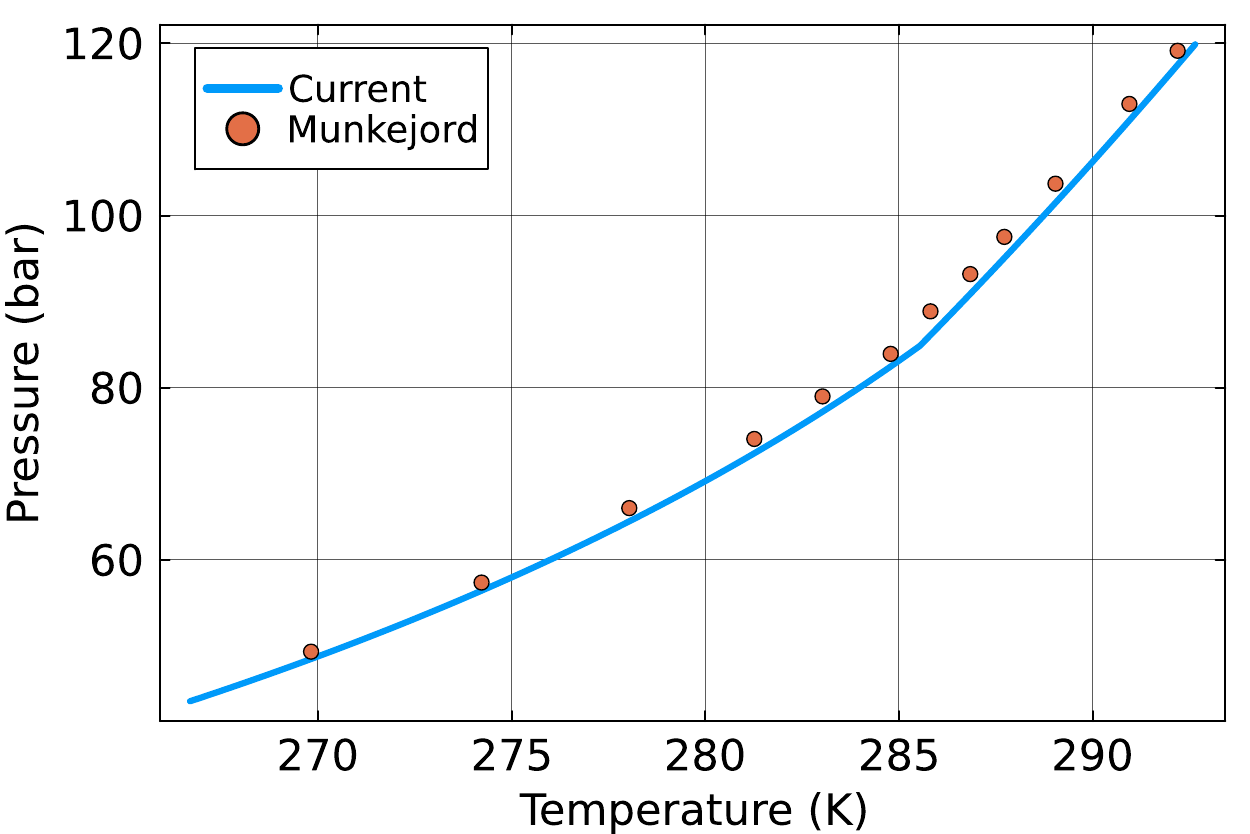}
        \caption{\coo and \nitrogen mixture \((90\%\--10\%)\) at \(t=0.3\)s}
        \label{fig.pipe.two_comps.800.case3a.HEMTrajectory.PT}
    \end{subfigure}   
    \begin{subfigure}[b]{0.48\textwidth}
        \includegraphics[width=\textwidth]{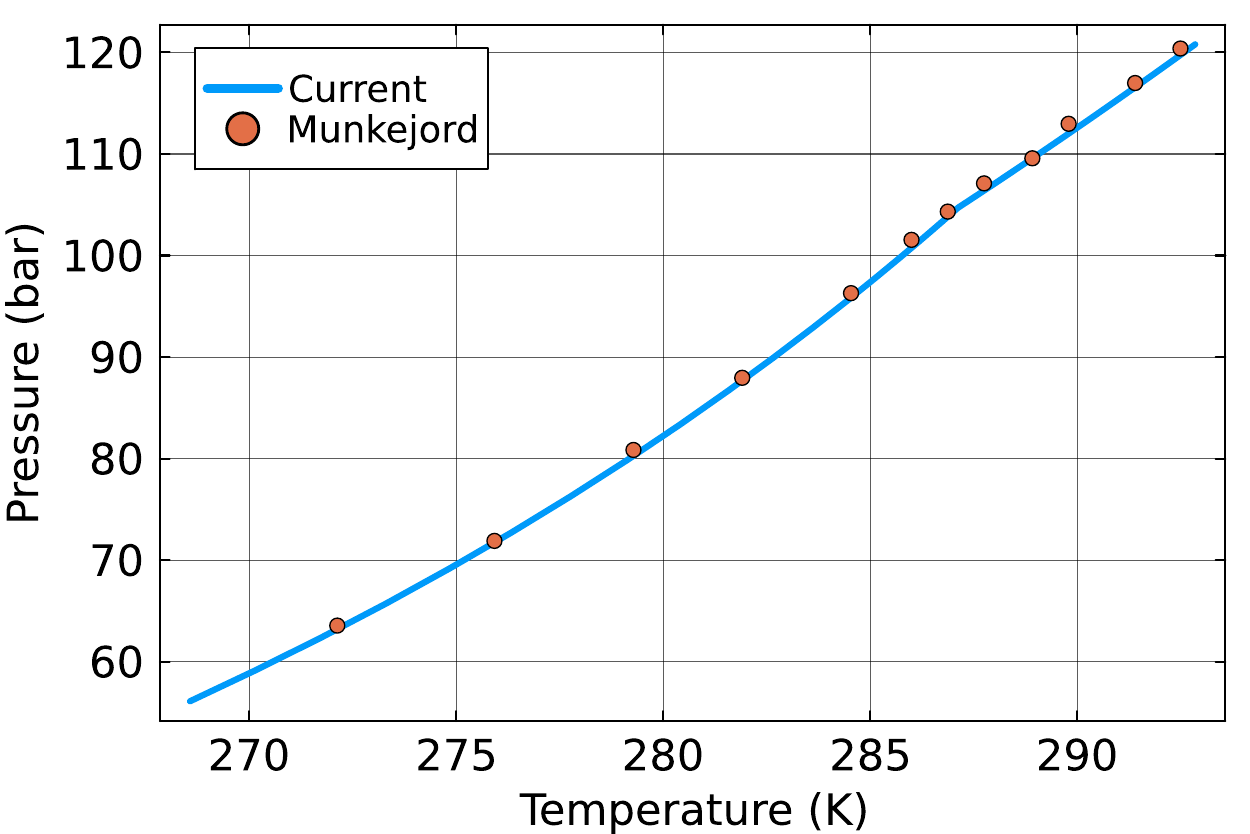}
        \caption{\coo and \nitrogen mixture \((80\%\--20\%)\) at \(t=0.3\)s }
        \label{fig.pipe.five_comps.800.case3a.HEMTrajectory.PT}
    \end{subfigure}  
    \begin{subfigure}[b]{0.48\textwidth}
        \includegraphics[width=\textwidth]{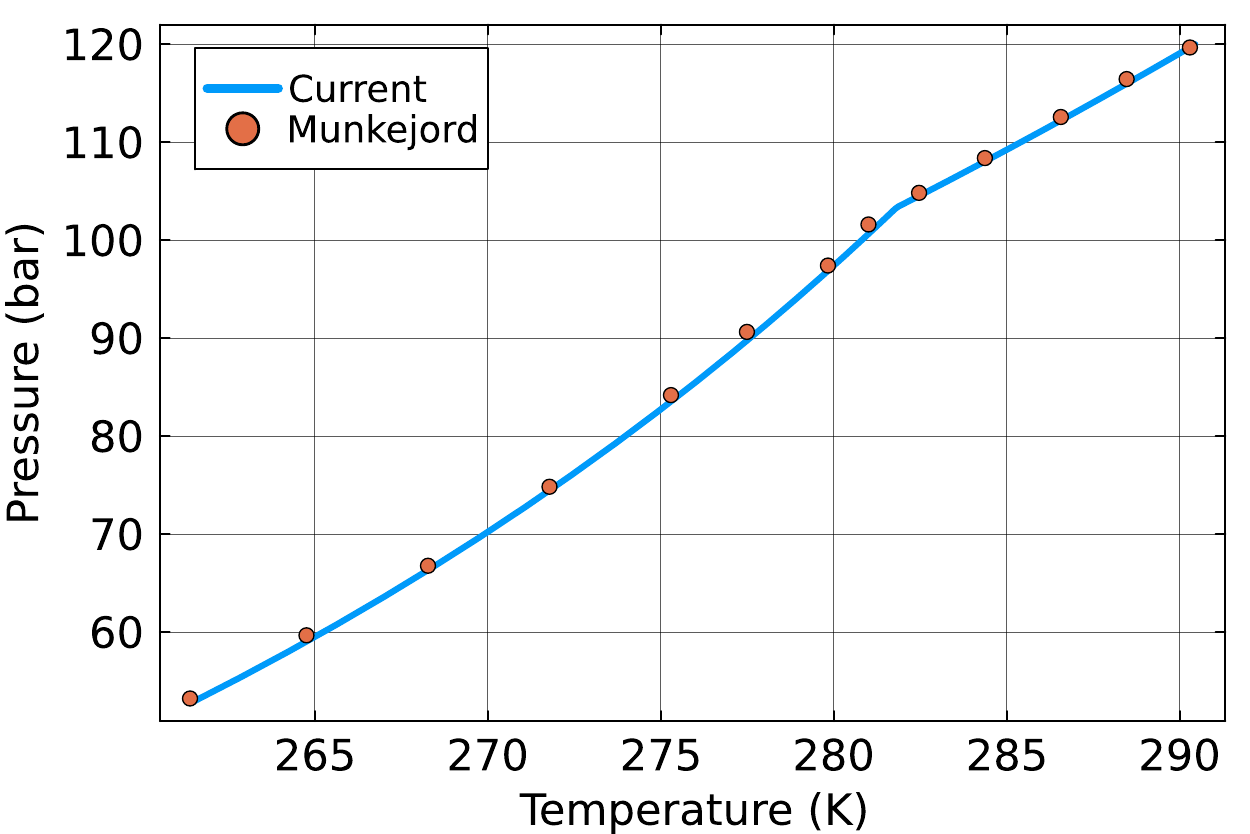}
        \caption{\coo and \nitrogen mixture \((70\%\--30\%)\) at \(t=0.3\)s}
        \label{fig.pipe.two_comps.800.case3c.HEMTrajectory.PT}
    \end{subfigure}
    \caption{Simulation Path in PT space}
    \label{fig.pipe.800.HEMTrajectory}
\end{figure}

The spatial convergence is discussed in \ref{app:spatial_convergence}. Here, we focus on the validation with the results reported by Munkejord \etal~\cite{munkejord_depressurization_2015} for  \ce{CO2}--rich mixtures, with specifications provided in \Cref{tab.pipe.shocktube.ic} and \Cref{tab.pipe.shocktube.composition}. It is important to highlight the differences between the setup considered here and that of Munkejord \etal~\cite{munkejord_depressurization_2015} as these distinctions may contribute to the discrepancies in the results. The main differences are: (i) we consider a shock-tube setup, whereas Munkejord \etal imposed a boundary condition, and (ii) the Peng–Robinson (PR) coefficients may differ, since they are not explicitly reported in Munkejord \etal \; Despite the differences in modeling choices, our results show good agreement with the literature, as can be seen in~\Cref{fig.pipe.800.HEMTrajectory}, where the HEM trajectory in the pressure--temperature (PT) plane is shown for the binary and the five-component mixtures. Notably, since we did not take into account the heat transfer and friction, the simulation path in the left section of the pipe corresponds to an isentropic process.

\begin{figure}[htbp]
    \centering
    \begin{subfigure}[b]{0.48\textwidth}
        \includegraphics[width=\textwidth]{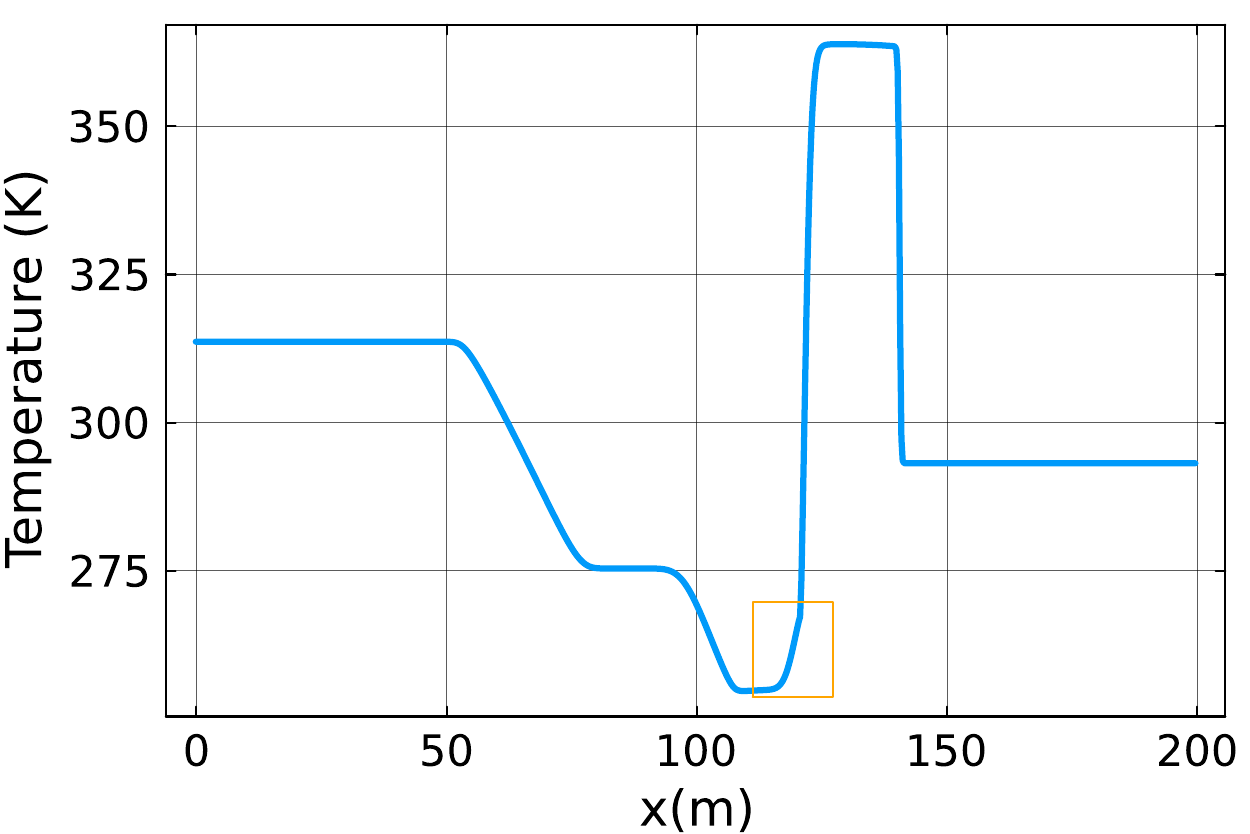}
        \caption{Temperature along the length of the pipe}
        \label{fig.pipe.two_comps.800.Tx}
    \end{subfigure}
    \hfill
    \begin{subfigure}[b]{0.48\textwidth}
        \includegraphics[width=\textwidth]{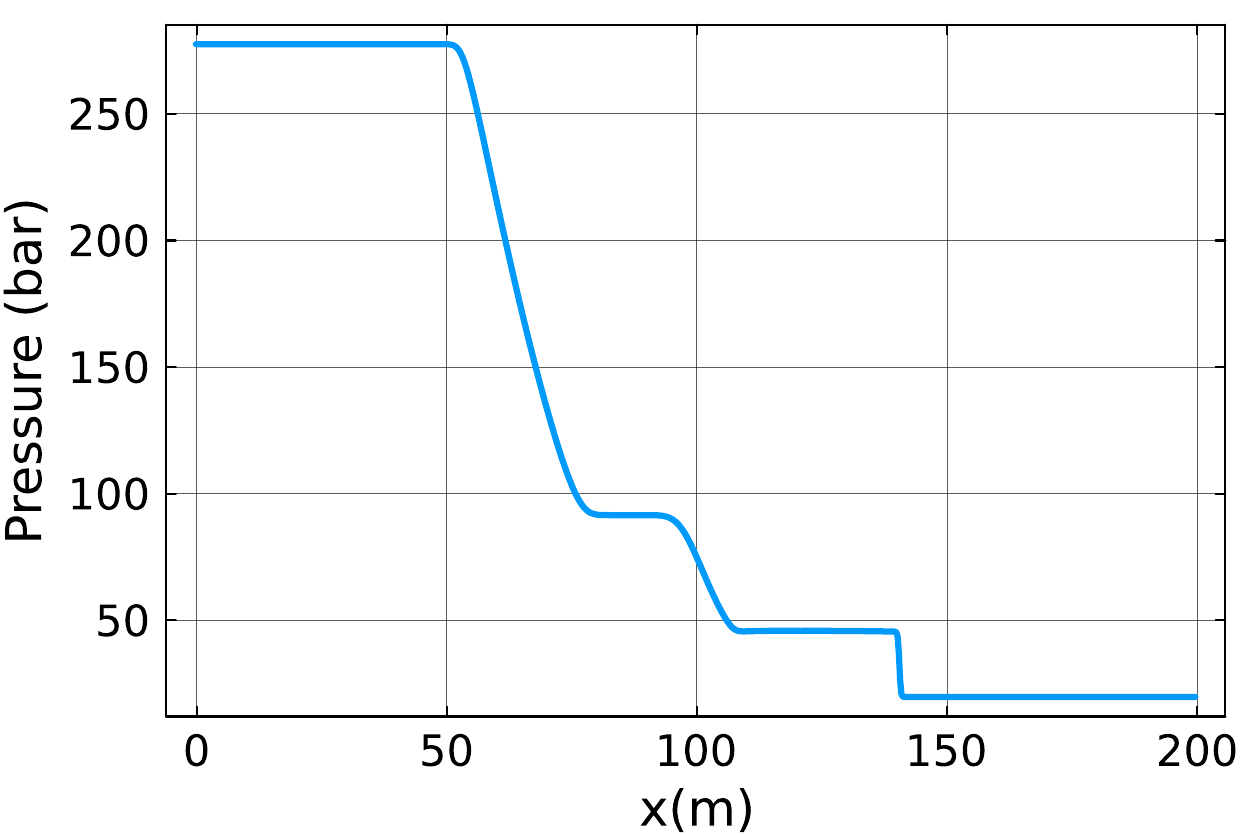}
        \caption{Pressure along the length of the pipe. }
        \label{fig.pipe.two_comps.800.Px}
    \end{subfigure}
    \caption{Shock tube results at $t = 0.1s$ with 800 cells for  \coo and \methane mixture.}
    \label{fig.pipe.two_comps.800.PT}
\end{figure}

\begin{figure}[htbp]
    \centering
    \begin{subfigure}[b]{0.48\textwidth}
        \includegraphics[width=\textwidth]{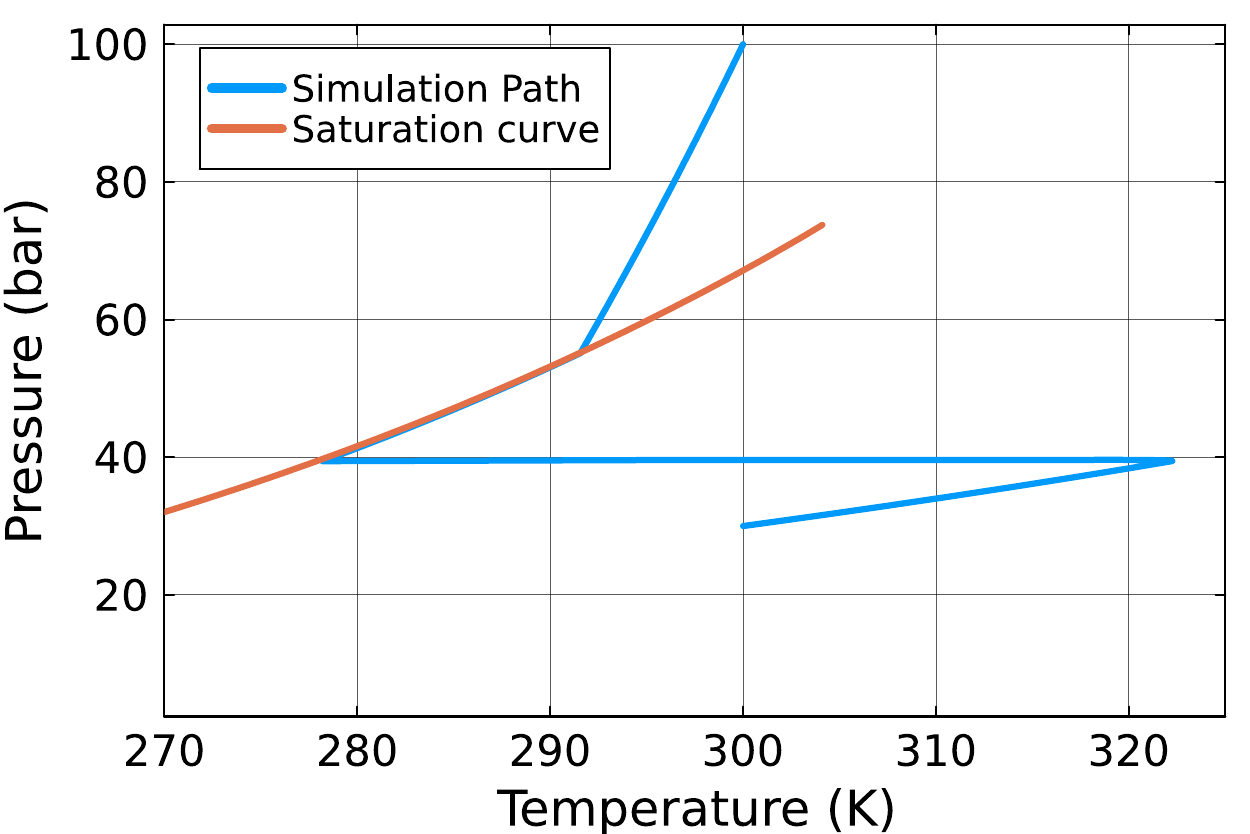}
        \caption{Pure \coo. }
        \label{fig.pipe.pure.800.PT}
    \end{subfigure}
    \hfill
    \begin{subfigure}[b]{0.48\textwidth}
        \includegraphics[width=\textwidth]{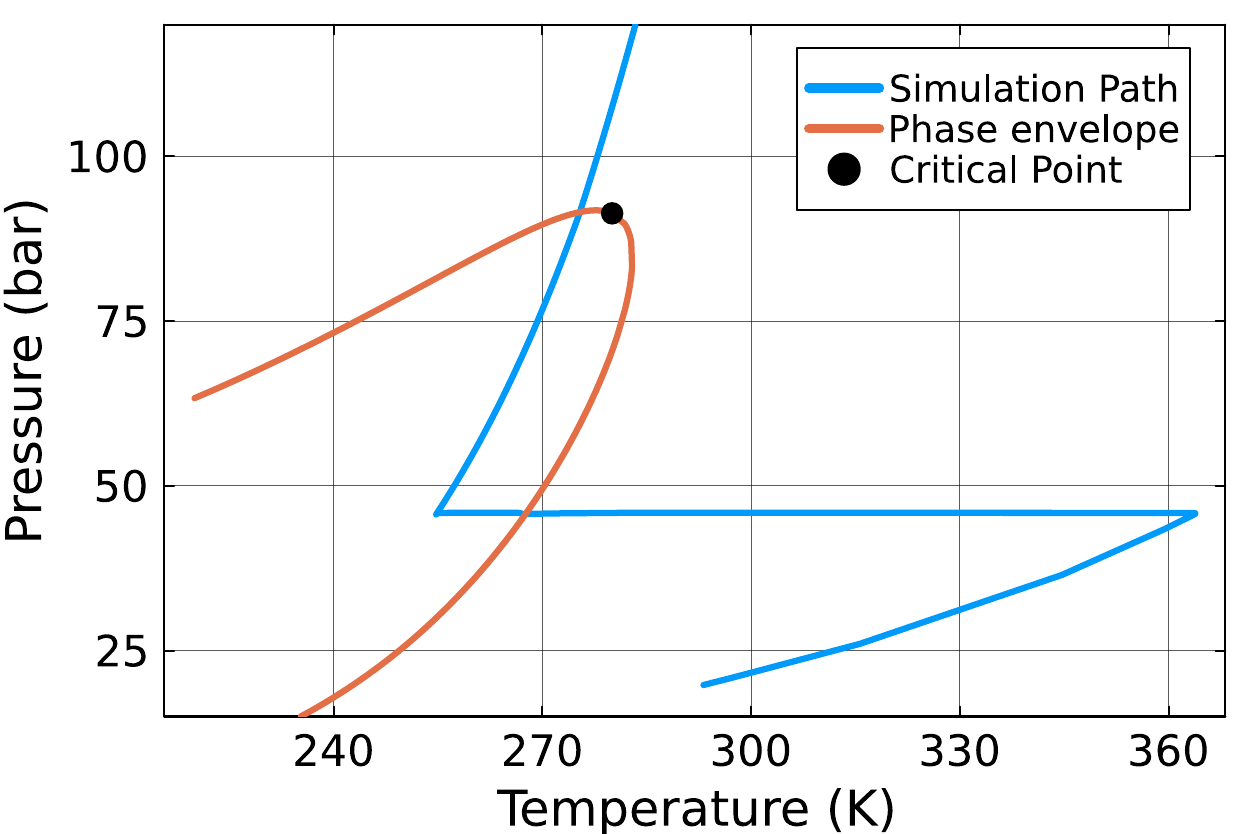}
        \caption{\coo and \methane mixture.}
        \label{fig.pipe.two_comps.800.PTVer3}
    \end{subfigure}
    \caption{Simulation path in PT-space}
    \label{fig.pipe.two_comps.800.simPath}
\end{figure}

\begin{figure}[H]
    \centering
    \begin{subfigure}[b]{0.48\textwidth}
        \includegraphics[width=\textwidth]{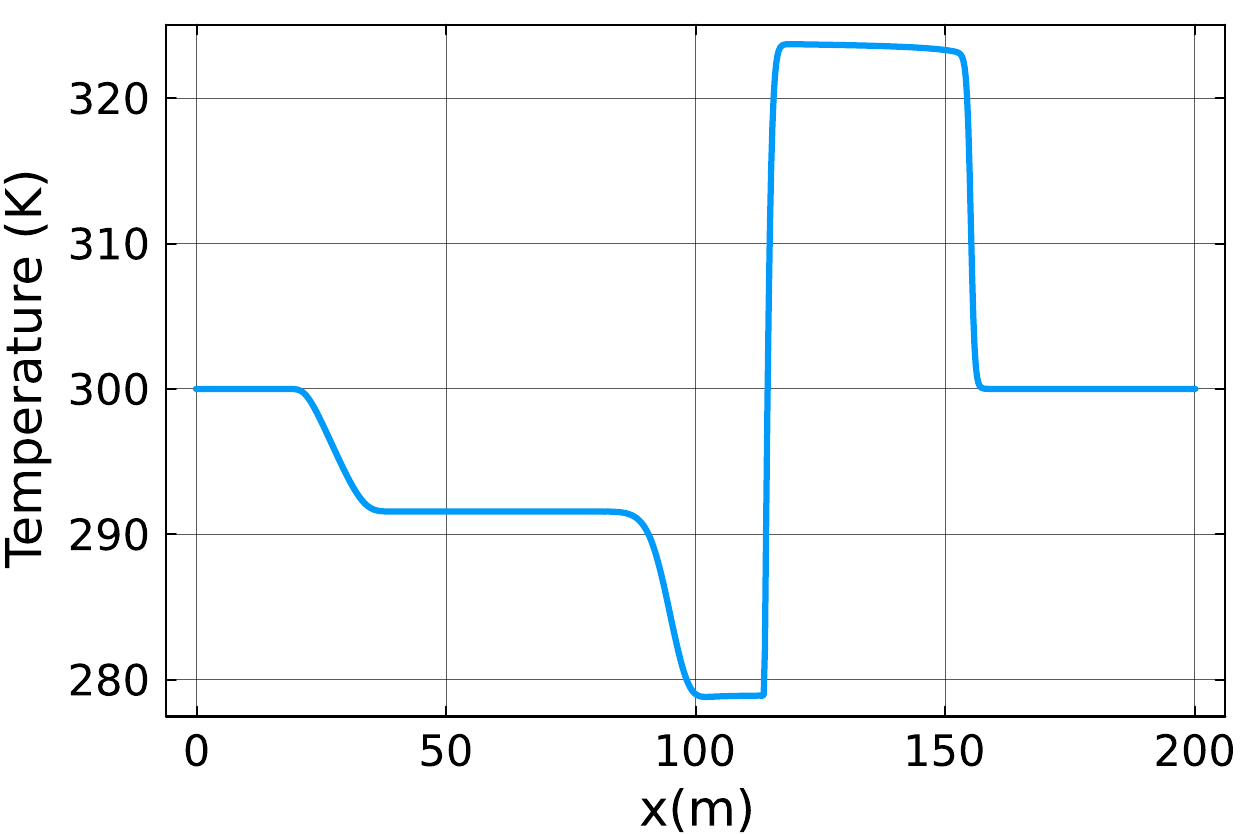} 
        \caption{Temperature along the length of the pipe}
        \label{fig.pipe.co2.800.Tx}
    \end{subfigure}
    \hfill
    \begin{subfigure}[b]{0.48\textwidth}
        \includegraphics[width=\textwidth]{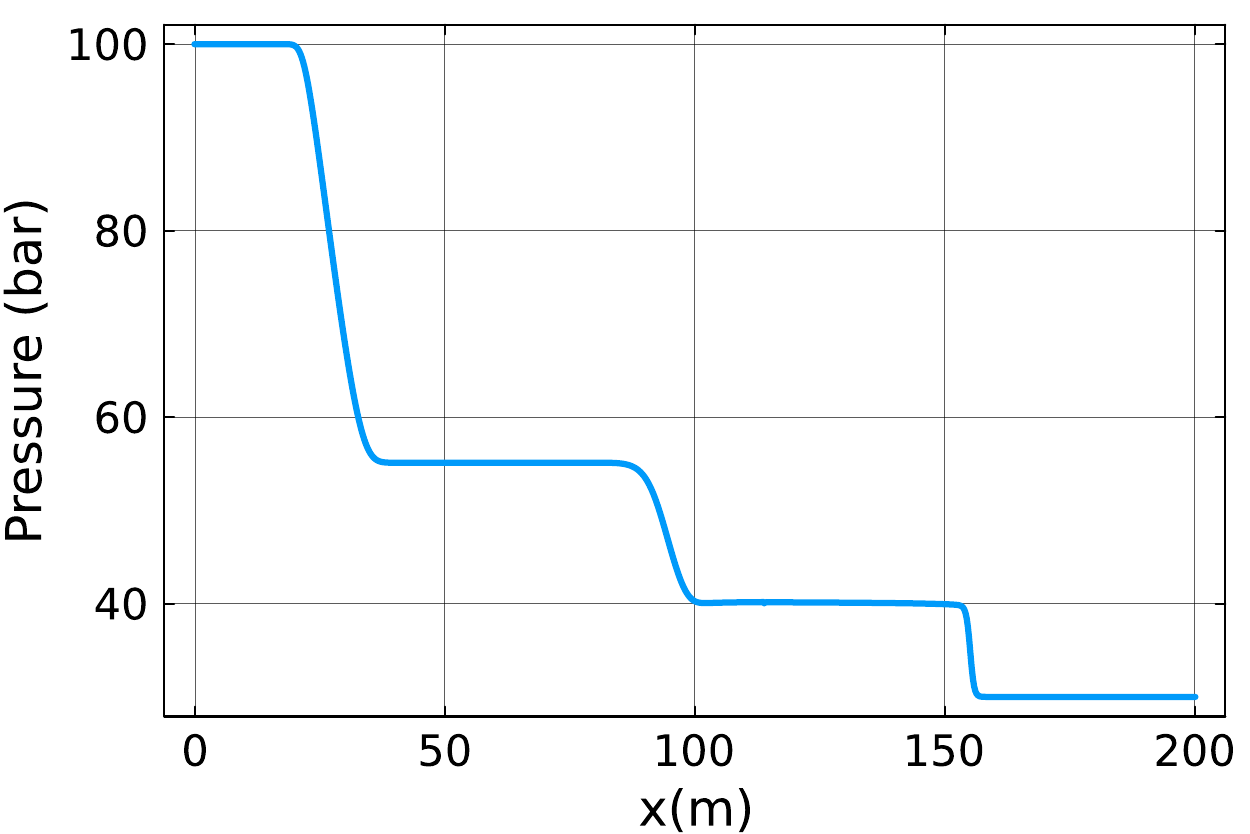}
        \caption{Pressure along the length of the pipe. }
        \label{fig.pipe.co2.800.Px}
    \end{subfigure}
    \caption{Shock tube results at $t = 0.2s$ with 800 grid cells for the single component(\coo).}
    \label{fig.pipe.co2.800}
\end{figure}

\Cref{fig.pipe.two_comps.800.PT} presents the results for pressure and temperature along the pipe length for a binary mixture of \ce{CO2} and \ce{CH4} at \(t=0.1s\). Around \(x \approx 120m\), a slight change in the slope in the temperature profile is observed as highlighted by an orange rectangle. This phenomenon is intrinsic to multicomponent mixtures and absent in single-component systems, as illustrated in the \Cref{fig.pipe.co2.800} where the evaporation wave ends abruptly into the contact wave. This difference can be explained by considering the topology of the phase diagram: for pure substances, the saturation points form a single curve in PT-space; see \Cref{fig.pipe.pure.800.PT}, where each pressure--temperature point can correspond to multiple values of phase fraction. For multicomponent mixtures, however, the two-phase region forms an envelope bounded by bubble and dew lines; see~\Cref{fig.pipe.two_comps.800.PTVer3}. Within this envelope, the iso-quality lines (where phase fraction remains constant) span the envelope from the bubble curve (vapor fraction $0$) to the dew curve (vapor fraction $1$). Consequently, each pressure--temperature condition inside the envelope corresponds to a unique value of vapor fraction.

\begin{figure}[htbp]
    \centering
    \begin{subfigure}[b]{0.48\textwidth}
        \includegraphics[width=\textwidth]{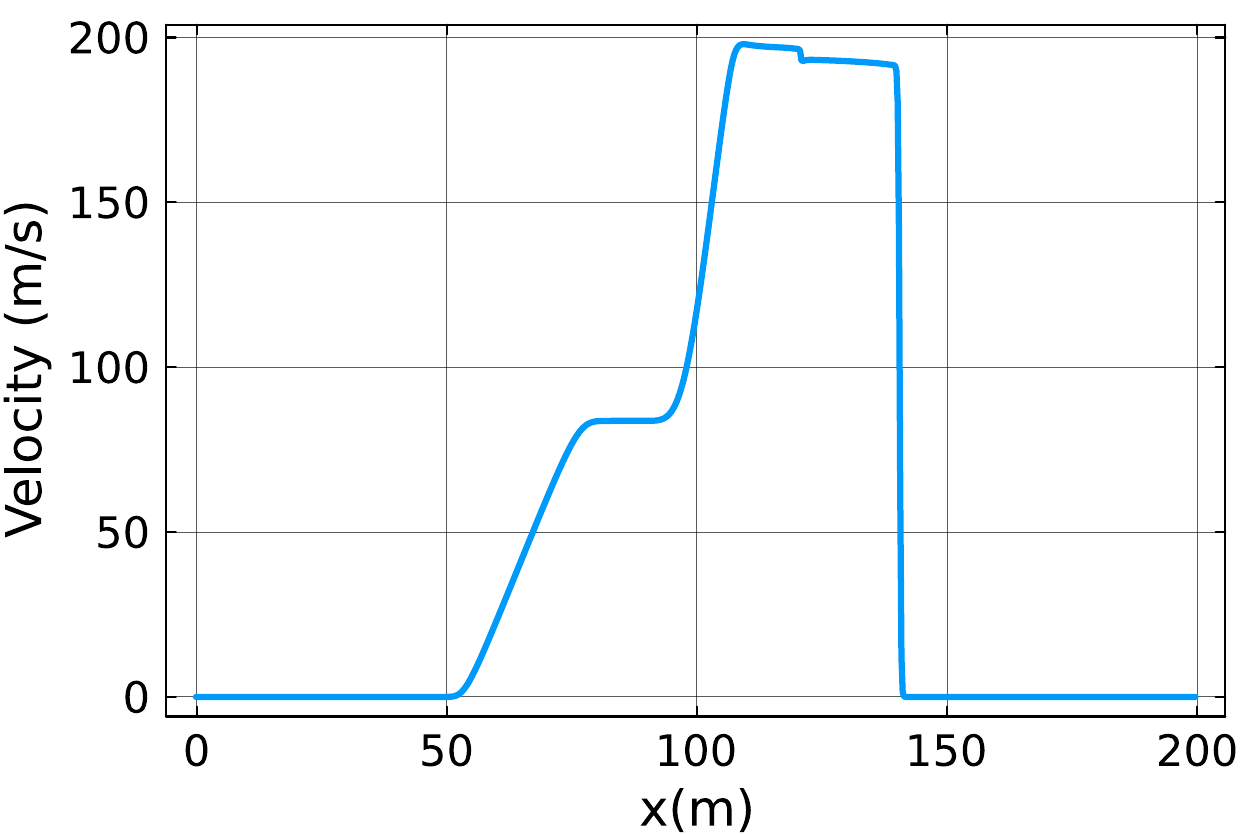}
        \caption{Fluid velocity along the length of the pipe}
        \label{fig.pipe.two_comps.800.Ux}
    \end{subfigure}
    \hfill
    \begin{subfigure}[b]{0.48\textwidth}
        \includegraphics[width=\textwidth]{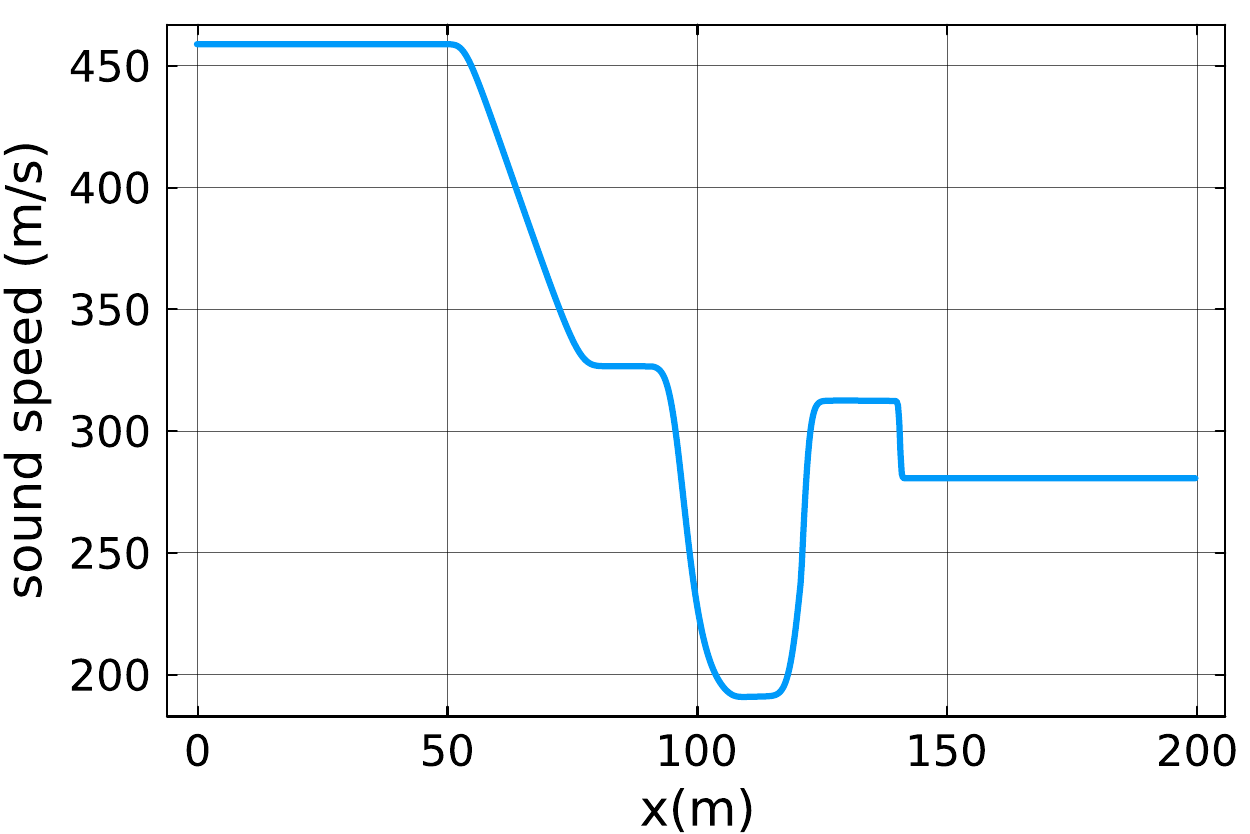}
        \caption{Sound speed along the length of the pipe. }
        \label{fig.pipe.two_comps.800.Ax}
    \end{subfigure}
    \caption{Shock tube results at $t = 0.1s$ with 800 cells for  \coo and \methane mixture.}
    \label{fig.pipe.two_comps.800.UA}
\end{figure}

\Cref{fig.pipe.two_comps.800.Ux} presents the velocity profile. Note that the fluid velocity changes across the contact wave, unlike the single-phase case. This has also been observed in \cite{foll_use_2019, kumar_new_2025}. \Cref{fig.pipe.two_comps.800.Ax} shows the spatial variation of the speed of sound along the pipeline. The speed of sound is an important component for calculations of wave speeds in approximate Riemann solvers (HLLC in this case); see \ref{HEM_HLLC}. A sharp drop in sound speed can be observed in the two-phase region when compared to the single-phase regions. This phenomenon is attributed to the increased compressibility of the fluid when both liquid and gaseous phases are present, and the gas bubbles dampen pressure disturbances. 

\begin{figure}[H]
    \centering
    \begin{subfigure}[b]{0.48\textwidth}
        \includegraphics[width=\textwidth]{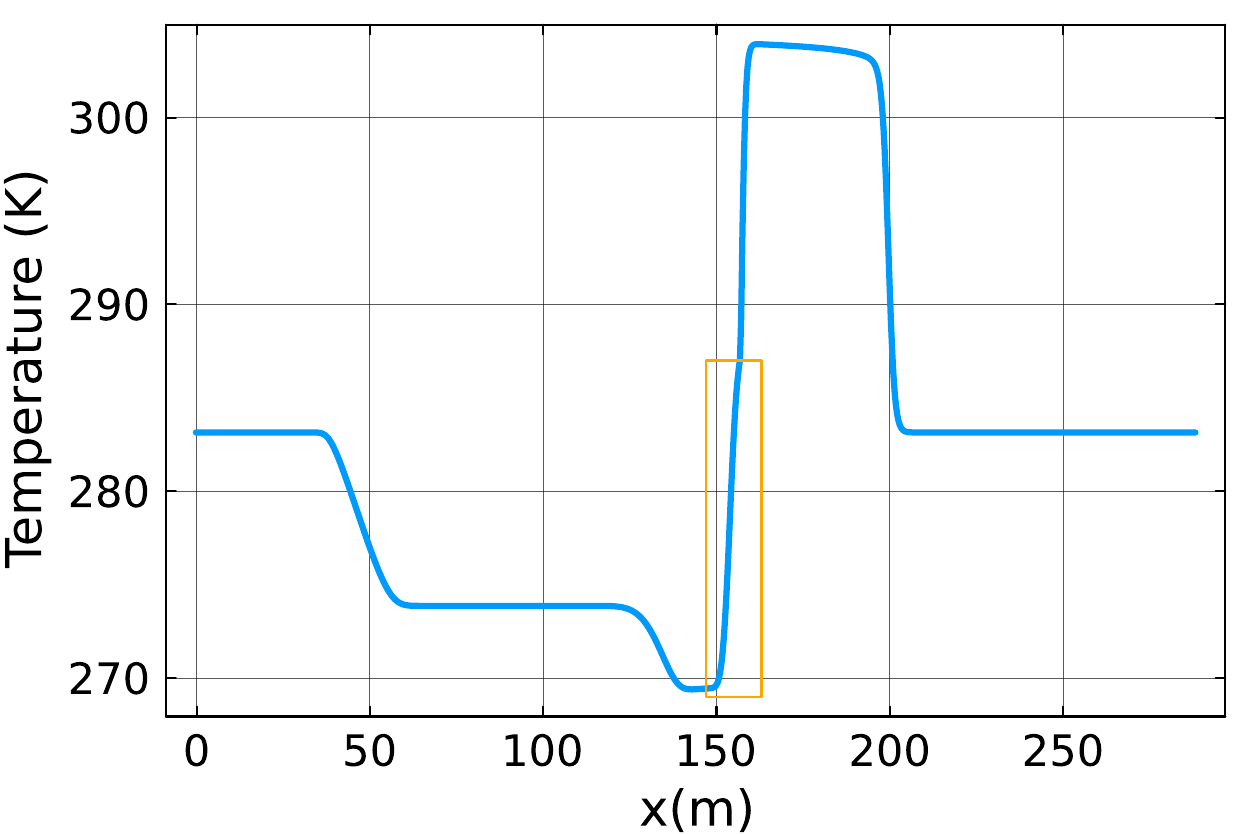} 
        \caption{Temperature along the length of the pipe}
        \label{fig.pipe.five_comps.800.Tx}
    \end{subfigure}
    \hfill
    \begin{subfigure}[b]{0.48\textwidth}
        \includegraphics[width=\textwidth]{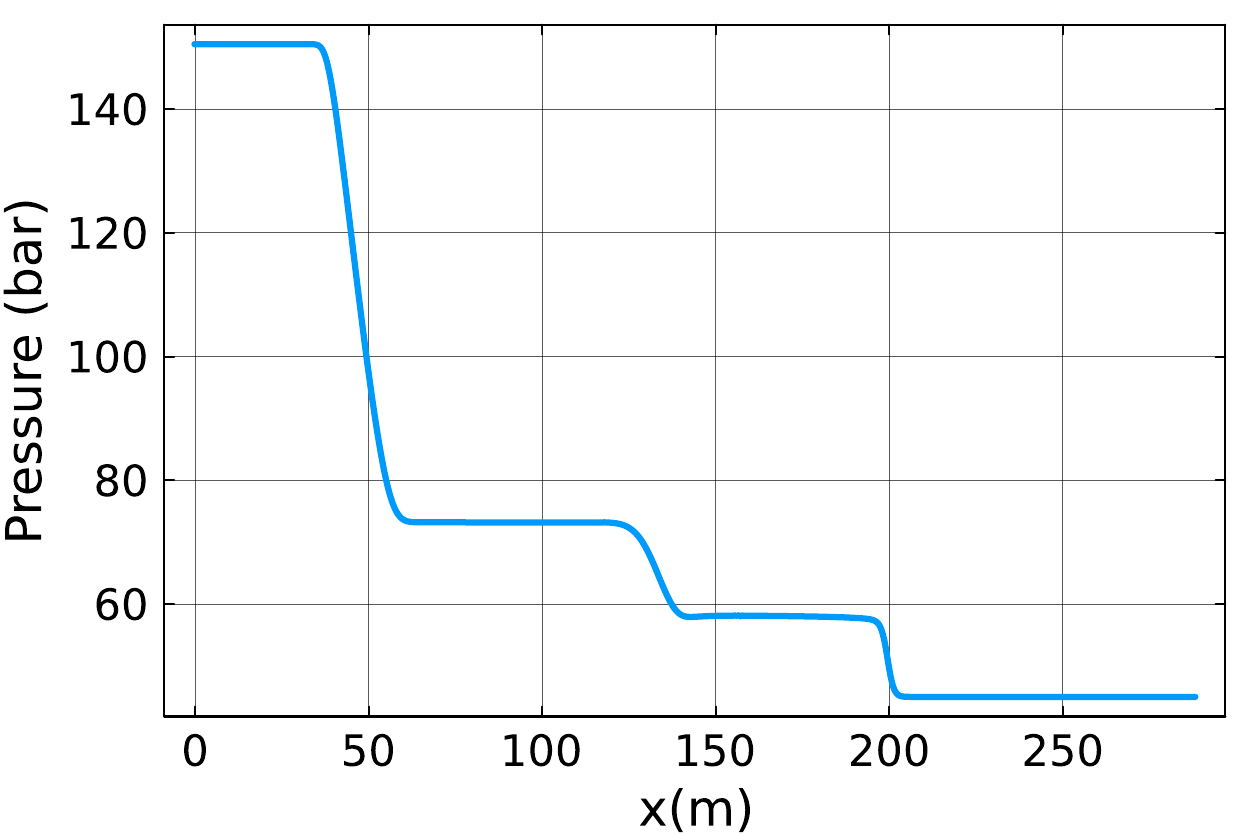}
        \caption{Pressure along the length of the pipe. }
        \label{fig.pipe.five_comps.800.Px}
    \end{subfigure}
    \caption{Shock tube results at $t = 0.22s$ with 800 cells for  five--component mixture.}
    \label{fig.pipe.five_comps.800}
\end{figure}

\begin{figure}[htbp]
    \centering
    \begin{subfigure}[b]{0.48\textwidth}
        \includegraphics[width=\textwidth]{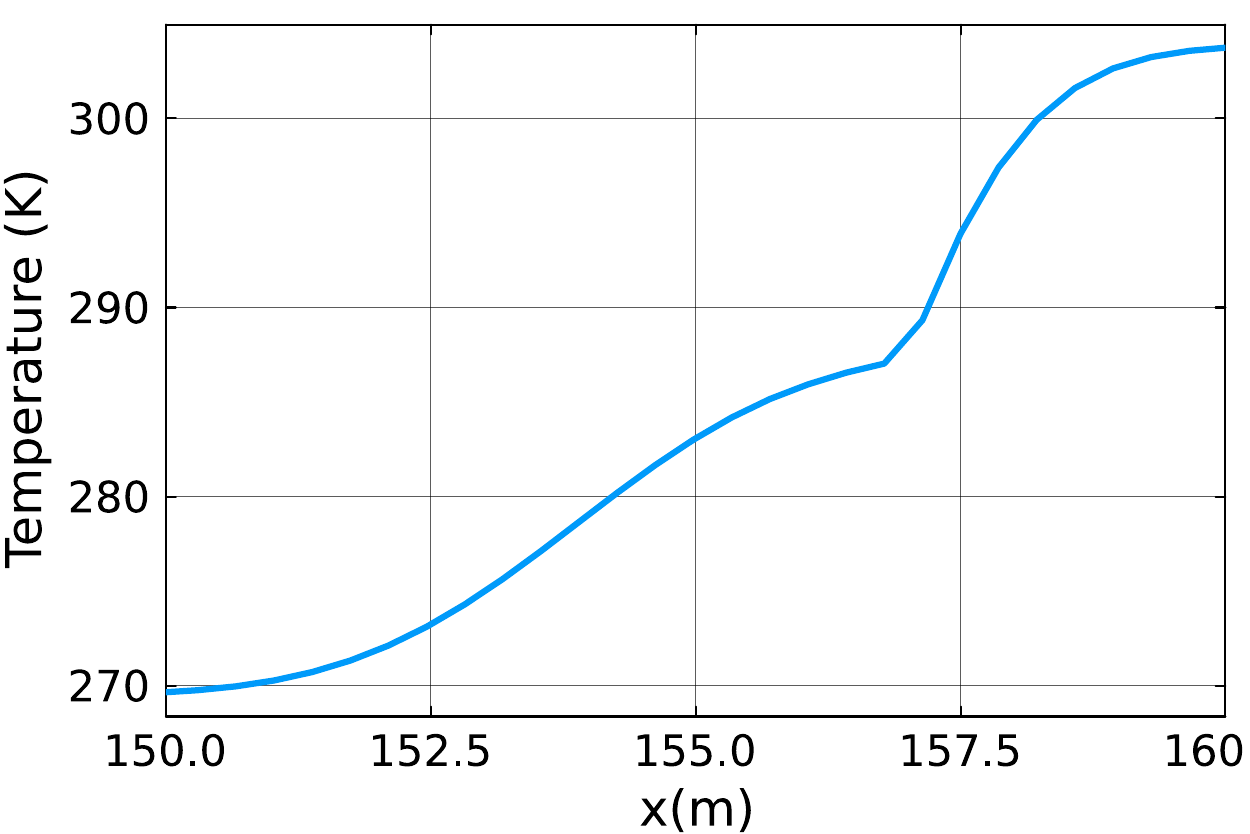} 
        \caption{Temperature along the length of the pipe}
        \label{fig.pipe.five_comps.800.Tx.zoomed}
    \end{subfigure}
    \hfill
    \begin{subfigure}[b]{0.48\textwidth}
        \includegraphics[width=\textwidth]{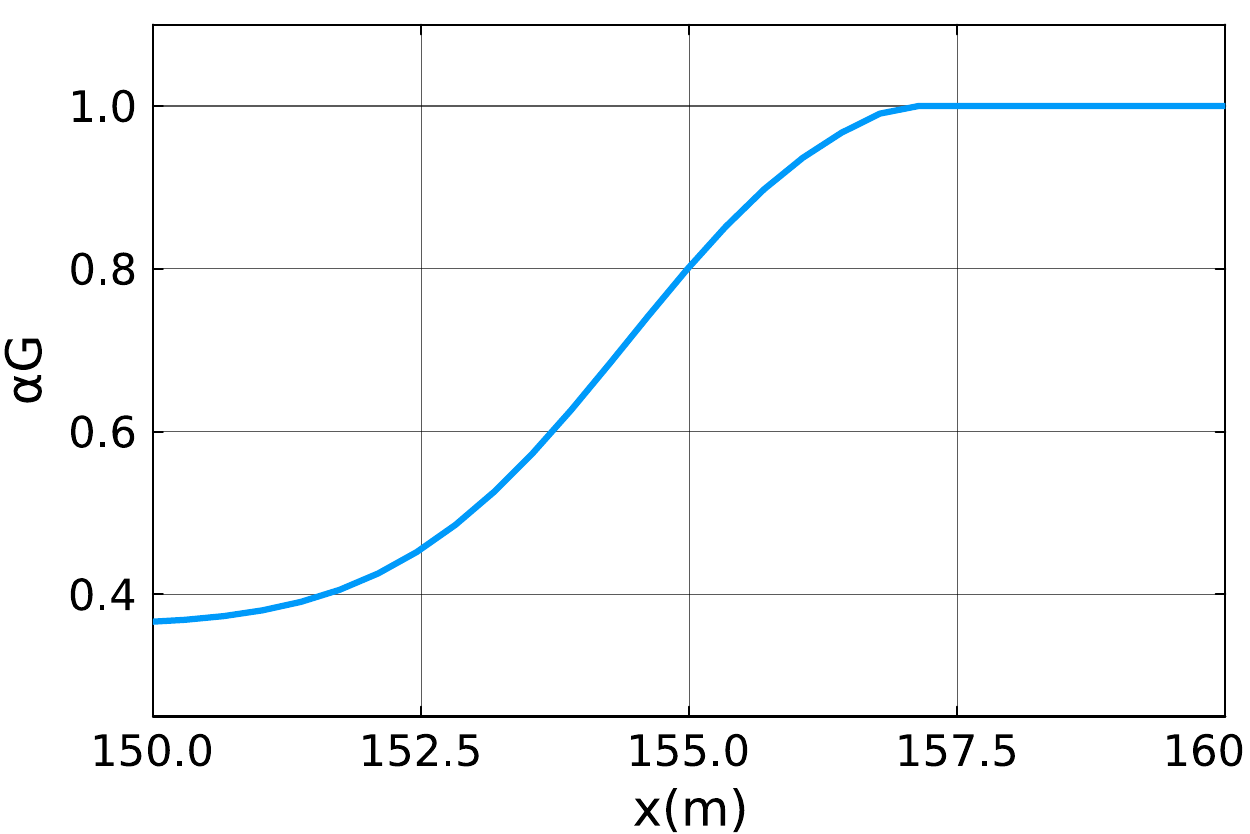}
        \caption{Vapor phase fraction along the length of the pipe. }
        \label{fig.pipe.five_comps.800.Ax.zoomed}
    \end{subfigure}
    \caption{Shock tube results at $t = 0.22$~s. 800 grid cells for the five-component mixture.}
    \label{fig.pipe.five_comps.800.zoomed}
\end{figure}

The results for the five-component mixture are presented in \Cref{fig.pipe.five_comps.800}. The overall wave structure is very similar to that of the binary case. The slight differences in wave amplitudes and slopes (highlighted with an orange rectangle) can be observed and arise due to differences in the initial conditions, composition, and the EOS coefficients. \Cref{fig.pipe.five_comps.800.Tx.zoomed} provides a magnified view of the sloping region. Notably, as is shown in \Cref{fig.pipe.five_comps.800.Px}, the pressure does not change across this region. Moreover, as illustrated in \Cref{fig.pipe.five_comps.800.Ax.zoomed}, this region corresponds to a transition from the two-phase to the single-phase regime. Finally, \Cref{fig.pipe.co2.800} presents the spatial pressure temperature (\Cref{fig.pipe.co2.800.Tx}) and (\Cref{fig.pipe.co2.800.Px}) profiles for the single-component \ce{CO2} case. 

To summarize, the proposed unified framework has been tested on multiple benchmark problems of varying complexity. It consistently shows a very good agreement with results from the literature, thereby demonstrating its robustness. It is important to emphasize that stability analysis and flash calculations are computationally demanding, and their corresponding routines are therefore natural candidates for future optimization.

\subsection{Wave structure}
For all three test cases, the solution exhibits a wave structure consistent with the typical wave patterns of a two-phase Riemann problem:
\begin{enumerate}
    \item \textbf{Left-moving rarefaction wave:} This smooth wave corresponds to the expansion of the fluid in the pipeline, leading to a drop in pressure and temperature across the wave. It is a genuinely non-linear wave in the context of the Riemann Problem~\cite{menikoff_riemann_1989}.
    \item \textbf{Evaporation wave:} This is also a smooth wave, across which liquid boiling occurs. Notably, this wave is absent in the Riemann problem for a single-phase case. Furthermore, for
    \begin{enumerate}
        \item \textit{Pure component}: The evaporation wave ends abruptly and transitions into the single-phase region with a contact discontinuity.
        \item \textit{Multicomponent}: The evaporation wave exhibits a sloping region that corresponds to varying values of vapor phase fractions in the phase envelope in \(PT\) space. Thereafter, it transitions into the single-phase region with a contact discontinuity.
    \end{enumerate}
   
    \item \textbf{Right-moving contact discontinuity:} This wave follows the shock wave and marks material separation. No fluid mixing occurs across this wave. Unlike in single-phase fluids, it is accompanied by a change in fluid velocity. %
    \item \textbf{Right-moving shock wave:} This wave corresponds to fluid compression, is accompanied by entropy production, and results in an increase in temperature and pressure.
\end{enumerate}

\section{Conclusion} \label{sec:conclusion}
In this paper, we have developed a unified framework that integrates fluid dynamics and thermodynamics for simulating the transport of multicomponent fluid in two-phase conditions. The \ce{CO2}--rich mixtures relevant to CCS applications were considered as test cases. We adopted the homogeneous equilibrium model (HEM) to describe the flow, and employed a Helmholtz free energy-based equation of state to model the thermodynamic properties. The key contribution is the development and testing of a tailored UVN-flash framework for dynamic pipeline transport models. The reformulation of the UVN-flash relies on the introduction of a new set of variables, namely, \(\rho^{\gas}_1, \dots, \rho^{\gas}_n, \alpha, T\). This choice aligns seamlessly with the inputs from the fluid dynamics solver, yielding better-scaled variables when compared to the standard UVN/TVN variables. Furthermore, we demonstrated that the critical points of the entropy function, when expressed in these variables, correspond to the classical thermodynamic equilibrium conditions. Additionally, we discussed a formulation of stability analysis for the UVN-flash which allows us to reliably detect single- and two-phase states. This is a crucial component for robust multiphase transient pipeline simulations.

The methodology was tested on a set of depressurization scenarios, including two tank problems (a binary and a five-component mixture) as well as pipeline cases for pure \ce{CO2}, four binary mixtures, and one five-component mixture. Numerical experiments demonstrate that the proposed framework provides a consistent and accurate description of the coupled thermodynamics and fluid dynamics problem, and is capable of capturing the extreme conditions relevant for safe pipeline design.

A flowchart of the algorithm and details of the temporal and spatial discretization of the resulting DAE system were also provided. Since thermodynamics computations are computationally intensive, exploring strategies for speeding up these calculations is an interesting research direction. Another promising research direction could be to investigate the application of SVN-flash in pipeline depressurization, as it could offer an efficient and accurate method for modeling the thermodynamics of isentropic flow.

\section*{CRediT author statement}

\textbf{Pardeep Kumar}:
Conceptualization,
Methodology,
Software,
Validation,
Analysis,
Writing -- Original Draft

\textbf{Patricio I. Rosen Esquivel}:
Project Administration,
Funding Acquisition,
Conceptualization,
Supervision,
Writing - Review \& Editing

\section*{Declaration of Generative AI and AI-assisted technologies in the writing process}

During the preparation of this work the authors used GitHub Copilot in order to
propose wordings and mathematical typesetting. After using this tool/service,
the authors reviewed and edited the content as needed. The authors take full
responsibility for the content of the publication.
\section*{Declaration of competing interest}

The authors declare that they have no known competing financial interests or
personal relationships that could have appeared to influence the work reported
in this paper.
\section*{Acknowledgments}  
This research was generously supported by Shell Projects and Technology, and we deeply appreciate their invaluable contribution.

\bibliographystyle{ieeetr}

\begin{thebibliography}{10}

\bibitem{noauthor_iea_2023}
``{IEA} (2023), {Energy} {Technology} {Perspectives}, {IEA}, {Paris}
  https://www.iea.org/reports/energy- technology-perspectives-2023, {License}:
  {CC} {BY} 4.0,'' tech. rep., 2023.

\bibitem{munkejord_thermo-_2010}
S.~T. Munkejord, J.~P. Jakobsen, A.~Austegard, and M.~J. M{\o}lnvik, ``Thermo-
  and fluid-dynamical modelling of two-phase multi-component carbon dioxide
  mixtures,'' {\em International Journal of Greenhouse Gas Control}, vol.~4,
  pp.~589--596, July 2010.

\bibitem{munkejord_depressurization_2015}
S.~T. Munkejord and M.~Hammer, ``Depressurization of {CO2}-rich mixtures in
  pipes: {Two}-phase flow modelling and comparison with experiments,'' {\em
  International Journal of Greenhouse Gas Control}, vol.~37, pp.~398--411, June
  2015.
\newblock Publisher: Elsevier Ltd.

\bibitem{munkejord_depressurization_2021}
S.~T. Munkejord, H.~Deng, A.~Austegard, M.~Hammer, A.~Aasen, and H.~L.
  Skarsv{\aa}g, ``Depressurization of {CO2}-{N2} and {CO2}-{He} in a pipe:
  {Experiments} and modelling of pressure and temperature dynamics,'' {\em
  International Journal of Greenhouse Gas Control}, vol.~109, p.~103361, July
  2021.

\bibitem{log_temperature_2025}
A.~M. Log, M.~Hammer, H.~Deng, A.~Austegard, and S.~T. Munkejord, ``Temperature
  response during rapid depressurization of {CO2} in a pipe: {Experiments} and
  fluid-dynamics modelling,'' {\em International Journal of Multiphase Flow},
  vol.~192, p.~105330, Nov. 2025.
\newblock Publisher: Elsevier BV.

\bibitem{drescher_experiments_2014}
M.~Drescher, K.~Varholm, S.~T. Munkejord, M.~Hammer, R.~Held, and
  G.~De~Koeijer, ``Experiments and modelling of two-phase transient flow during
  pipeline depressurization of {CO2} with various {N2} compositions,'' {\em
  Energy Procedia}, vol.~63, pp.~2448--2457, 2014.

\bibitem{cosham_decompression_2012}
A.~Cosham, D.~G. Jones, and J.~Barnett, ``The decompression behaviour of carbon
  dioxide in the dense phase,'' tech. rep., 2012.

\bibitem{botros_measuring_2013}
K.~Botros, E.~Hippert, and P.~Craidy, ``Measuring decompression wave speed in
  {CO2} mixtures by a shock tube,'' {\em Pipelines Int}, vol.~16, pp.~22--28,
  Jan. 2013.

\bibitem{bruce_stewart_two-phase_1984}
H.~Bruce~Stewart and B.~Wendroff, ``Two-phase flow: {Models} and methods,''
  {\em Journal of Computational Physics}, vol.~56, pp.~363--409, Dec. 1984.

\bibitem{munkejord_co2_2016}
S.~T. Munkejord, M.~Hammer, and S.~W. L{\o}vseth, ``{CO2} transport: {Data} and
  models – {A} review,'' {\em Applied Energy}, vol.~169, pp.~499--523, May
  2016.

\bibitem{toumi_approximate_1999}
I.~Toumi, A.~Kumbaro, and H.~Pailler, ``Approximate {Riemann} solvers and flux
  vector splitting schemes for two-phase flow,'' Tech. Rep. CEA-R-5849(E),
  1999.

\bibitem{saurel_multiphase_1999}
R.~Saurel and R.~Abgrall, ``A {Multiphase} {Godunov} {Method} for
  {Compressible} {Multifluid} and {Multiphase} {Flows},'' tech. rep., 1999.
\newblock Publication Title: Journal of Computational Physics Volume: 150.

\bibitem{saurel_relaxation-projection_2007}
R.~Saurel, E.~Franquet, E.~Daniel, and O.~Le~Metayer, ``A relaxation-projection
  method for compressible flows. {Part} {I}: {The} numerical equation of state
  for the {Euler} equations,'' {\em Journal of Computational Physics},
  vol.~223, pp.~822--845, May 2007.
\newblock Publisher: Academic Press Inc.

\bibitem{abgrall_computations_2001}
R.~Abgrall and S.~Karni, ``Computations of {Compressible} {Multifluids},'' {\em
  Journal of Computational Physics}, vol.~169, pp.~594--623, May 2001.
\newblock Publisher: Academic Press Inc.

\bibitem{kapila_two-phase_2001}
A.~K. Kapila, R.~Menikoff, J.~B. Bdzil, S.~F. Son, and D.~S. Stewart,
  ``Two-phase modeling of deflagration-to-detonation transition in granular
  materials: {Reduced} equations,'' {\em Physics of Fluids}, vol.~13, no.~10,
  pp.~3002--3024, 2001.
\newblock Publisher: American Institute of Physics Inc.

\bibitem{pelanti_mixture-energy-consistent_2014}
M.~Pelanti and K.~M. Shyue, ``A mixture-energy-consistent six-equation
  two-phase numerical model for fluids with interfaces, cavitation and
  evaporation waves,'' {\em Journal of Computational Physics}, vol.~259,
  pp.~331--357, Feb. 2014.
\newblock Publisher: Academic Press Inc.

\bibitem{michelsen_isothermal_1981}
M.~L. Michelsen, ``The isothermal flash problem. {Part} {II}. {Phase}-split
  calculation.,'' {\em Fluid Phase Equilibria}, 1981.

\bibitem{michelsen_isothermal_1982}
M.~L. Michelsen, ``The isothermal flash problem. {Part} {I}. {Stability}.,''
  {\em Fluid Phase Equilibria}, 1982.

\bibitem{michelsen_thermodynamic_2007}
M.~L. Michelsen and J.~M. Mollerup, {\em Thermodynamic models : fundamentals \&
  computational aspects}.
\newblock Tie-Line Publications, 2007.

\bibitem{holyst_thermodynamics_2012}
R.~Holyst and A.~Poniewierski, {\em Thermodynamics for chemists, physicists and
  engineers}.
\newblock Springer Netherlands, Jan. 2012.
\newblock Publication Title: Thermodynamics for Chemists, Physicists and
  Engineers.

\bibitem{nichita_calculation_2007}
D.~V. Nichita, D.~Broseta, and F.~Montel, ``Calculation of convergence
  pressure/temperature and stability test limit loci of mixtures with cubic
  equations of state,'' {\em Fluid Phase Equilibria}, vol.~261, pp.~176--184,
  Dec. 2007.

\bibitem{nichita_isochoric_2009}
D.~V. Nichita, J.-C. de~Hemptinne, and S.~Gomez, ``Isochoric {Phase}
  {Stability} {Testing} for {Hydrocarbon} {Mixtures},'' {\em Petroleum Science
  and Technology}, vol.~27, pp.~2177--2191, Oct. 2009.

\bibitem{nichita_rapid_2013}
D.~V. Nichita and C.~F. Leibovici, ``A rapid and robust method for solving the
  {Rachford}–{Rice} equation using convex transformations,'' {\em Fluid Phase
  Equilibria}, vol.~353, pp.~38--49, Sept. 2013.

\bibitem{nichita_robustness_2023}
D.~V. Nichita, ``Robustness and efficiency of phase stability testing at {VTN}
  and {UVN} conditions,'' {\em Fluid Phase Equilibria}, vol.~564, p.~113624,
  Jan. 2023.

\bibitem{elshahomi_decompression_2015}
A.~Elshahomi, C.~Lu, G.~Michal, X.~Liu, A.~Godbole, and P.~Venton,
  ``Decompression wave speed in {CO2} mixtures: {CFD} modelling with the
  {GERG}-2008 equation of state,'' {\em Applied Energy}, vol.~140, pp.~20--32,
  Feb. 2015.

\bibitem{munkejord_depressurization_2020}
S.~T. Munkejord, A.~Austegard, H.~Deng, M.~Hammer, H.~G. Stang, and S.~W.
  L{\o}vseth, ``Depressurization of {CO2} in a pipe: {High}-resolution pressure
  and temperature data and comparison with model predictions,'' {\em Energy},
  vol.~211, Nov. 2020.
\newblock Publisher: Elsevier Ltd.

\bibitem{hammer_method_2013}
M.~Hammer, A.~Ervik, and S.~T. Munkejord, ``Method {Using} a
  {Density}–{Energy} {State} {Function} with a {Reference} {Equation} of
  {State} for {Fluid}-{Dynamics} {Simulation} of {Vapor}–{Liquid}–{Solid}
  {Carbon} {Dioxide},'' {\em Industrial \& Engineering Chemistry Research},
  vol.~52, pp.~9965--9978, July 2013.

\bibitem{morin_two-fluid_2013}
A.~Morin, T.~Fl{\aa}tten, and F.~Fl{\aa}tten, ``A two-fluid four-equation model
  with instantaneous thermodynamical equilibrium,'' tech. rep., 2013.

\bibitem{aursand_spinodal_2017}
P.~Aursand, M.~A. Gjennestad, E.~Aursand, M.~Hammer, and {\o}.~Wilhelmsen,
  ``The spinodal of single- and multi-component fluids and its role in the
  development of modern equations of state,'' {\em Fluid Phase Equilibria},
  vol.~436, pp.~98--112, Mar. 2017.

\bibitem{lachet_equilibrium_2012}
V.~Lachet, B.~Creton, T.~De~Bruin, E.~Bourasseau, N.~Desbiens,
  {\o}.~Wilhelmsen, and M.~Hammer, ``Equilibrium and transport properties of
  {CO2}+{N2O} and {CO2}+{NO} mixtures: {Molecular} simulation and equation of
  state modelling study,'' {\em Fluid Phase Equilibria}, vol.~322-323,
  pp.~66--78, May 2012.

\bibitem{wilhelmsen_thermodynamic_2017}
O.~Wilhelmsen, A.~Aasen, G.~Skaugen, P.~Aursand, A.~Austegard, E.~Aursand,
  M.~A. Gjennestad, H.~Lund, G.~Linga, and M.~Hammer, ``Thermodynamic
  {Modeling} with {Equations} of {State}: {Present} {Challenges} with
  {Established} {Methods},'' {\em Industrial \& Engineering Chemistry
  Research}, vol.~56, pp.~3503--3515, Apr. 2017.

\bibitem{ishii_thermo-fluid_2011}
M.~Ishii and T.~Hibiki, {\em Thermo-fluid dynamics of two-phase flow}.
\newblock New York: Springer, 2nd ed~ed., 2011.
\newblock OCLC: ocn690084123.

\bibitem{castier_dynamic_2010}
M.~Castier, ``Dynamic simulation of fluids in vessels via entropy
  maximization,'' {\em Journal of Industrial and Engineering Chemistry},
  vol.~16, pp.~122--129, Jan. 2010.

\bibitem{arendsen_dynamic_2009}
A.~R.~J. Arendsen and G.~F. Versteeg, ``Dynamic thermodynamics with internal
  energy, volume, and amount of moles as states: {Application} to {Liquefied}
  {Gas} {Tank},'' {\em Industrial \& Engineering Chemistry Research}, vol.~48,
  pp.~3167--3176, Mar. 2009.

\bibitem{saha_isoenergetic-isochoric_1997}
S.~Saha and J.~J. Carroll, ``The isoenergetic-isochoric flash,'' {\em Fluid
  Phase Equilibria}, vol.~138, pp.~23--41, Nov. 1997.

\bibitem{lima_differential-algebraic_2008}
E.~R. Lima, M.~Castier, and E.~C. Biscaia, ``Differential-{Algebraic}
  {Approach} to {Dynamic} {Simulations} of {Flash} {Drums} with {Rigorous}
  {Evaluation} of {Physical} {Properties},'' {\em Oil \& Gas Science and
  Technology - Revue de l'IFP}, vol.~63, pp.~677--686, Sept. 2008.

\bibitem{qiu_multiphase_2014}
L.~Qiu, Y.~Wang, and R.~D. Reitz, ``Multiphase dynamic flash simulations using
  entropy maximization and application to compressible flow with phase
  change,'' {\em AIChE Journal}, vol.~60, pp.~3013--3024, Aug. 2014.

\bibitem{giljarhus_solution_2012}
K.~E.~T. Giljarhus, S.~T. Munkejord, and G.~Skaugen, ``Solution of the
  {Span}-{Wagner} equation of state using a density-energy state function for
  fluid-dynamic simulation of carbon dioxide,'' in {\em Industrial and
  {Engineering} {Chemistry} {Research}}, vol.~51, pp.~1006--1014, Journal of
  Fluids and Structures, Jan. 2012.
\newblock Issue: 2 ISSN: 08885885.

\bibitem{kumar_new_2025}
P.~Kumar, B.~Sanderse, P.~I.~R. Esquivel, and R.~Henkes, ``A new temperature
  evolution equation that enforces thermodynamic vapour–liquid equilibrium in
  multiphase flows - application to {CO2} modelling,'' {\em Computers \&
  Fluids}, vol.~289, p.~106524, Mar. 2025.
\newblock Publisher: Elsevier BV.

\bibitem{peng_new_1976}
D.-Y. Peng and D.~B. Robinson, ``A new two-constant equation of state,'' {\em
  Industrial \& Engineering Chemistry Fundamentals}, vol.~15, pp.~59--64, Feb.
  1976.
\newblock Number: 1.

\bibitem{nichita_new_2018}
D.~V. Nichita, ``New unconstrained minimization methods for robust flash
  calculations at temperature, volume and moles specifications,'' {\em Fluid
  Phase Equilibria}, vol.~466, pp.~31--47, June 2018.

\bibitem{smejkal_phase_2017}
T.~Smejkal and J.~Mikyška, ``Phase stability testing and phase equilibrium
  calculation at specified internal energy, volume, and moles,'' {\em Fluid
  Phase Equilibria}, vol.~431, pp.~82--96, Jan. 2017.

\bibitem{kumar_solving_2026}
P.~Kumar and P.~I.~R. Esquivel, ``Solving the {UVN}-flash problem in
  {TVN}-space,'' {\em Fluid Phase Equilibria}, vol.~599, p.~114528, Jan. 2026.
\newblock Publisher: Elsevier BV.

\bibitem{smejkal_smejkal_2021}
T.~Smejkal, {\em Smejkal {Thesis}}.
\newblock PhD thesis, 2021.

\bibitem{mikyska_investigation_2012}
J.~Mikyška and A.~Firoozabadi, ``Investigation of mixture stability at given
  volume, temperature, and number of moles,'' {\em Fluid Phase Equilibria},
  vol.~321, pp.~1--9, May 2012.

\bibitem{castier_solution_2009}
M.~Castier, ``Solution of the isochoric–isoenergetic flash problem by direct
  entropy maximization,'' {\em Fluid Phase Equilibria}, vol.~276, pp.~7--17,
  Feb. 2009.

\bibitem{hairer_numerical_1989}
E.~Hairer, C.~Lubich, and M.~Roche, {\em The numerical solution of
  differential-algebraic systems by {Runge}-{Kutta} methods}.
\newblock No.~1409 in Lecture notes in mathematics, Berlin Heidelberg:
  Springer, 1989.

\bibitem{foll_use_2019}
F.~Föll, T.~Hitz, C.~Müller, C.-D. Munz, and M.~Dumbser, ``On the use of
  tabulated equations of state for multi-phase simulations in the homogeneous
  equilibrium limit,'' {\em Shock Waves}, vol.~29, pp.~769--793, July 2019.
\newblock Publisher: Springer Science and Business Media LLC.

\bibitem{menikoff_riemann_1989}
R.~Menikoff and B.~J. Plohr, ``The {Riemann} problem for fluid flow of real
  materials,'' {\em Reviews of Modern Physics}, vol.~61, pp.~75--130, Jan.
  1989.

\bibitem{toro_riemann_2009}
E.~F. Toro, {\em Riemann {Solvers} and {Numerical} {Methods} for {Fluid}
  {Dynamics}}.
\newblock 2009.

\bibitem{span_new_1996}
R.~Span and W.~Wagner, ``A {New} {Equation} of {State} for {Carbon} {Dioxide}
  {Covering} the {Fluid} {Region} from the {Triple}-{Point} {Temperature} to
  1100 {K} at {Pressures} up to 800 {MPa},'' {\em Journal of Physical and
  Chemical Reference Data}, vol.~25, pp.~1509--1596, Nov. 1996.

\bibitem{drew_theory_1999}
D.~A. Drew and S.~L. Passman, {\em Theory of {Multicomponent} {Fluids}}.
\newblock Applied {Mathematical} {Sciences}, New York, NY: Springer New York,
  1999.
\newblock ISSN: 0066-5452, 2196-968X.

\bibitem{poling_properties_2004}
B.~Poling, J.~Prausnitz, and J.~O'Connel, {\em The {Properties} of {Gases} and
  {Liquids}}.
\newblock 2004.

\end{thebibliography}

\appendix

\section{Spatial Convergence} \label{app:spatial_convergence}
\begin{figure}[H]
    \centering
        \includegraphics[width=0.5\textwidth]{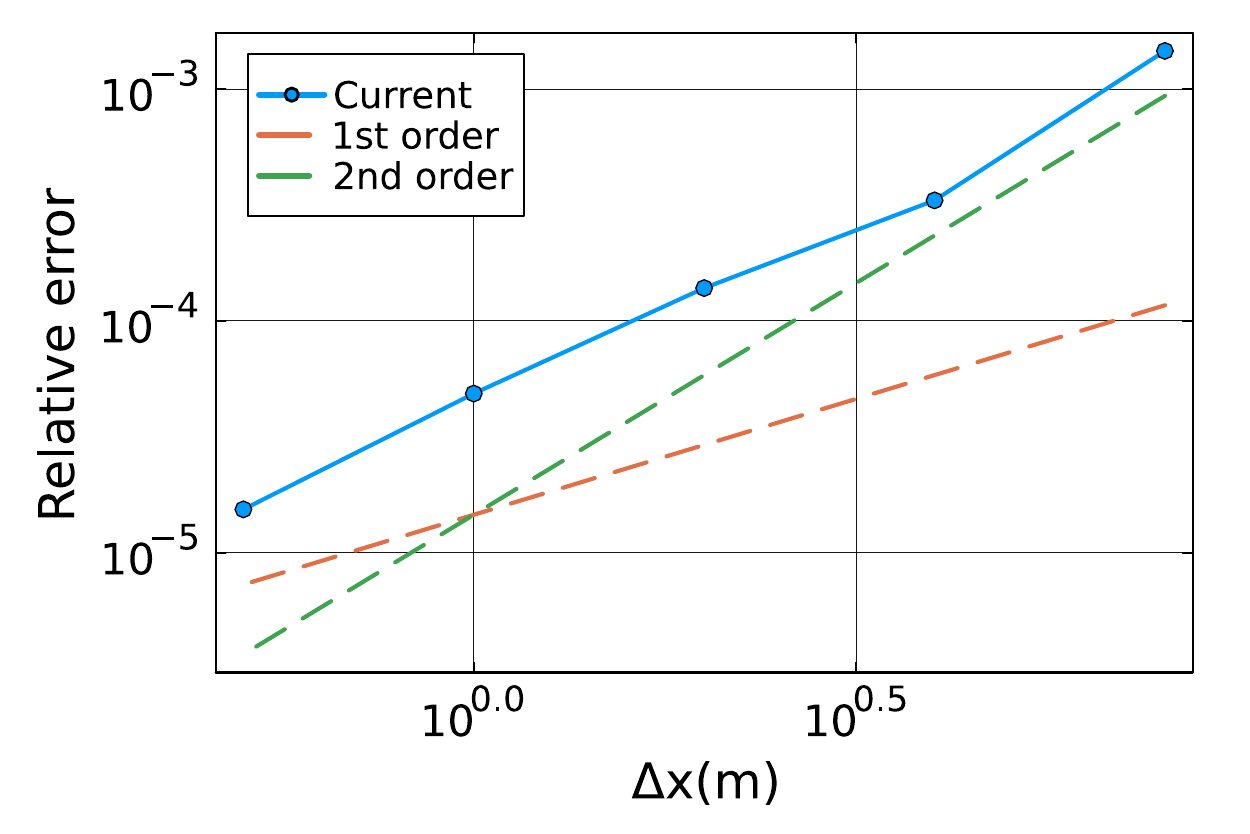} 
    \caption{Pipeline depressurization: Spatial convergence along the pipe at \(t = 0.07s\).}
    \label{fig.pipe.co2.convergence.spatial.T}
\end{figure}

To evaluate the spatial convergence in the pipeline simulations, we consider the binary mixture case. A high-resolution run on a mesh of 6400 cells is used as the reference solution. The \(L_1\) error in temperature is reported in \Cref{fig.pipe.co2.convergence.spatial.T}, where the results are plotted on a log-log scale. We observe increased accuracy as the mesh is refined. The observed convergence rate lies between first and second-order accuracy.

\section{Spatial discretization of HEM model} \label{HEM_HLLC}

To compute the numerical flux $\hat{\EulerFlux}_{i+\frac{1}{2}}$, we employ the HLLC (Harten–-Lax–-van Leer–-Contact) scheme which is an approximate Riemann solver \cite{toro_riemann_2009}. This method exploits the wave structure of the Riemann problem to compute fluxes at the interfaces between adjacent computational cells. For single-phase compressible flow, the Riemann solution typically consists of three distinct wave types: rarefaction waves, shock waves, and a contact discontinuity. In the context of two-phase flow, however, an additional discontinuity, referred to as the evaporation wave, may arise due to phase transition phenomena. The HLLC scheme approximates the fluxes associated with each of these waves by enforcing the Rankine-Hugoniot condition. A complete treatment of the HLLC scheme for Euler's equations can be found in Toro \cite{toro_riemann_2009}. Below, we present the final expressions for the numerical fluxes at the interface $i+\frac{1}{2}$:

\begin{align}
    \hat{\EulerFlux}_{i+1/2} &= 
    \begin{cases}
        \hllcF_L, & \text{if } 0 \leq S_L, \\
        \hllcF^{\star}_{L}, & \text{if } S_L \leq 0 \leq S^{\star}, \\
        \hllcF^{\star}_{R}, & \text{if } S^{\star} \leq 0 \leq S_R, \\
        \hllcF_R, & \text{if } 0 \geq S_R .
    \end{cases} \\
    \hllcF^{\star}_{L} &= \hllcF_L + S_L (U^{\star}_{L} - U_L), \\ 
    \hllcF^{\star}_{R} &= \hllcF_R + S_R (U^{\star}_{R} - U_R), \\
    S_L &= \min(u_L - a_L, u_R - a_R), \\
    S_R &= \max(u_L + a_L, u_R + a_R), \\
    S^{\star} &= \frac{p_R - p_L + \rho_L u_L (S_L - u_L) - \rho_R u_R (S_R - u_R)}{\rho_L (S_L - u_L) - \rho_R (S_R - u_R)}, \\
    U^{\star}_{K} &= \rho_K \left( 
    \frac{S_K - u_K}{S_K - S^{\star}} \right) \begin{bmatrix}
        1 \\
        S^{\star} \\
        E_K + (S^{\star} - u_K) \left [ S^{\star} + \frac{p_K}{\rho_K(S_K - u_K)}\right]
    \end{bmatrix}.
\end{align}
Here, $K = L$(left state) or $K = R$(right state), $U_K \approx \cU(x_{K},t)$, $\hllcF_{K} = \EulerFlux(U_{K}), a_{K}$ denotes the local speed of sound computed using the equation of state \cite{span_new_1996}. The wave speeds $S_L$ and $S_R$ represent the velocities of the left and right-propagating waves, respectively, and $S^{\star}$ is the speed of the contact wave. The accuracy of the HLLC scheme is strongly influenced by the quality of the wave speeds($S_L$ and $S_R$) estimates. Here, we have presented one such method for estimating these wave speeds. For a broader discussion of wave speed approximation, the interested reader is referred to Toro \cite{toro_riemann_2009}.

\section{Wood's Speed of Sound for an $n$-Component, Two-Phase Mixture}
\label{sec:woods_pr}

Consider an $n$-component mixture in mechanical equilibrium between a gas ($\gas$) and a liquid ($\liq$) phase. Let the vapor volume fraction be $\phaseFrac \in [0,1]$. Denote the sound speeds by $a_\gas, a_\liq$ and mass densities by $\rho_\gas, \rho_\liq$ of the vapor and liquid phase, respectively. Wood's mixture relation~\cite{drew_theory_1999} then reads 
\begin{equation}
\frac{1}{\rho_m a_m^2}
=
\frac{\phaseFrac}{\rho_\gas a_\gas^2}
+
\frac{1-\phaseFrac}{\rho_\liq a_\liq^2},
\qquad
\rho_m = \phaseFrac \rho_\gas + (1-\phaseFrac)\rho_\liq ,
\label{eq:woods-mixture}
\end{equation}
where \(\rho_m\) is the overall mixture density.

\subsection*{\textbf{Phase densities from $N$ and $V$}}
Let $N_i^{\gas}$ and $N_i^{\liq}$ be the moles of component $i$ in vapor and liquid, respectively, and $M_{w,i}$ the molar mass. With phase volumes $V^{\gas}$ and $V^{\liq}$, the phasic component mass densities can be written as:
\begin{align}
\rho_i^{\gas} &= \frac{N_i^{\gas} M_{w,i}}{V^{\gas}}, \quad \rho_\gas = \sum_{i=1}^n \rho_i^{\gas}, \\
\rho_i^{\liq} &= \frac{N_i^{\liq} M_{w,i}}{V^{\liq}}, \quad \rho_\liq = \sum_{i=1}^n \rho_i^{\liq}.
\end{align}

\subsection*{\textbf{Phasic speed of sound}}
Using the thermodynamic identity for the isentropic speed of sound for each phase
\begin{equation}
a^2 = \left( \frac{\partial p}{\partial \rho} \right)_{S}
= \frac{V}{\rho} \left[ -\frac{\partial p}{\partial V} + \frac{T}{C_v} \left( \frac{\partial p}{\partial T} \right)^2 \right],
\label{eq:speed-of-sound}
\end{equation}
where \(C_v\) is the isochoric heat-capacity and \(\rho\) is the phasic density (\(\rho_{\gas/\liq}\)).

\section{Peng--Robinson Equation of State} \label{app:PR_EOS}
We employ the Peng--Robinson equation of state (EOS) \cite{smejkal_phase_2017}, which is formulated as follows:
\begin{equation}
    P(T, V, N_1, \ldots, N_n) = \frac{NRT}{V - B} - \frac{a(T)N^2}{V^2 + 2BV - B^2}, \label{eq:peng_robinson}
\end{equation}
where \(T\) is the temperature, \(V\) is the volume, \(N_i\) represents the number of moles of component \(i\) in the system, \(R\) is the universal gas constant and \(N\) is the total number of moles in the system. The parameter \(a(T)\) characterizes the attractive intermolecular forces, while the effective co-volume $B$ is defined in terms of the parameter $b$, which characterizes the repulsive interactions, as:

\begin{subequations}
\begin{equation} \label{eq:B_parameter}
    B = b N,
\end{equation}
\begin{equation}
    a = \sum_{i=1}^{n} \sum_{j=1}^{n} x_i x_j a_{ij}, \label{eq:a_parameter}
\end{equation}
\begin{equation} 
    a_{ij} = (1 - \delta_{ij}) \sqrt{a_i a_j}, \label{eq:a_ij}
\end{equation}
\begin{equation}
    a_i(T) = 0.45724 \frac{R^2 T_{\text{crit},i}^2}{P_{\text{crit},i}} \left[1 + m_i \left(1 - \sqrt{T_{r,i}} \right)\right]^2, \label{eq:a_i}
\end{equation}
\begin{equation}
    b = \sum_{i=1}^{n} x_i b_i, \label{eq:b_parameter}
\end{equation}
\begin{equation}
    b_i = 0.0778 \frac{R T_{\text{crit},i}}{P_{\text{crit},i}}, \label{eq:b_i}
\end{equation}
\end{subequations}
where \(x_i = N_i/N\) is the mole fraction of component \(i\),  \(T_{\text{crit},i}\), \(P_{\text{crit},i}\) and \(T_{r, i} = T/T_{\text{crit},i}\) are the critical temperature, critical pressure and the reduced temperature of component \(i\), and \(\delta_{ij}\) is the binary interaction parameter between component \(i\) and \(j\). The parameter \(m_i\) accounts for the acentric factor \(\omega_i\) as:
\begin{equation}
    m_i = \begin{cases} 
    0.37464 + 1.54226 \omega_i - 0.26992 \omega_i^2, & \omega_i < 0.5, \\
    0.3796 + 1.485 \omega_i - 0.1644 \omega_i^2 + 0.01667 \omega_i^3, & \omega_i \geq 0.5.
    \end{cases} \label{eq:m_i}
\end{equation}
\newline
The residual internal energy, \(U\), in the context of the Peng-Robinson EOS is expressed as follows.
\begin{align} 
    U(T, V, N_1, \ldots, N_n) &= N \frac{T \partial_T(a) - a} {2 \sqrt{2}b} \ln \left[ \frac{V + \delta_1 B}{V + \delta_2  B} \right] \nonumber \\
    &- NR(T - T_0) + \sum_{i=1}^{n} N_i \int_{T_0}^{T} c_{p,i}^{\mathrm{ig}}(\xi) \, d\xi + N u_0, \label{eq:internal_energy}
\end{align}
where \(\partial_T(a)\) is the temperature derivative of \(a(T)\), \(T_0\) is a reference temperature, \(\alpha_{ik}\) are empirical constants, \( \delta_1 = 1 + \sqrt{2}\) and \( \delta_2 = 1-\sqrt{2} \). The residual entropy, \(S\), is given as 
\begin{align}
    S(T, V, N_1, \ldots, N_n) &= NR \ln \left[ \frac{V - B}{V} \right] + N\frac{\partial_T(a)}{2 \sqrt{2}b} \ln \left[ \frac{V + \delta_1 B}{V + \delta_2  B} \right] \nonumber \\
    &+ R \sum_{i=1}^{n} N_i \ln \frac{V P_0}{N_i RT} + \sum_{i=1}^{n} N_i \int_{T_0}^{T} \frac{c_{p,i}^{\text{ig}}(\xi)}{\xi} d\xi, \label{eq:entropy}
\end{align}
where \(c_{p,i}^{\text{ig}}(T)\) is the ideal gas heat capacity of component \(i\) and \(P_0\) is a reference pressure. The heat capacity \(c_{p,i}^{\text{ig}}(T)\) can be written as:
\begin{equation}
    c_{p,i}^{\text{ig}}(T) = \sum_{k=0}^{3} \alpha_{ik} T^k. \label{eq:heat_capacity}
\end{equation}
Now, we can simplify the integral in \eqref{eq:internal_energy} as \[ \int_{T_0}^{T} c_{p,i}^{\mathrm{ig}}(\xi) \, d\xi = \sum_{k=0}^{3} \alpha_{ik} \frac{T^{k+1} - T_0^{k+1}}{k+1}, \]
and the integral in \eqref{eq:entropy} as \[ \int_{T_0}^{T} \frac{c_{p,i}^{\text{ig}}(\xi)}{\xi} d\xi, = \alpha_{i0} \ln \left( \frac{T}{T_0} \right) + \sum_{k=1}^{3} \alpha_{ik} \frac{T^{k} - T_0^{k}}{k}. \]
The coefficients \(\alpha_0, \alpha_1, \alpha_2, \alpha_3\)  for the fluids considered in this work are listed in Table~\ref{tab:cp_1_9}, while the parameters of the Peng–Robinson equation of state are summarized in Table~\ref{tab:pr_params}. It is important to note that the arguments of logarithmic terms must remain positive in Equations \eqref{eq:internal_energy} and \eqref{eq:entropy}. If this condition is violated, the current step should be rejected or appropriately truncated to maintain physical consistency.
The reference state is specified at \( T_0 = 298.15\,\mathrm{K} \) and \( P_0 = 1\,\mathrm{bar} \), where the molar internal energy is defined as
    \[
    u_0 = u(T_0, P_0) = h(T_0, P_0) - RT_0 = -RT_0 = -2478.95687512\,\mathrm{J\,mol^{-1}}.
    \]
This definition ensures that the molar enthalpy of the ideal gas at the reference conditions is zero \cite{smejkal_phase_2017}, i.e., \( h(T_0, P_0) = 0 \). Furthermore, the molar entropy of each pure component as an ideal gas is also set to zero at this state, \( s_i^{\text{ideal}}(T_0, P_0) = 0 \).

\begin{table}[H]
\centering
\caption{Correlation coefficients \( c_p^{\text{ig}} \)~\cite{smejkal_phase_2017, poling_properties_2004}.}
\label{tab:cp_1_9}
\begin{tabular}{llccc}
\toprule
Component & \(\alpha_0\) & \(\alpha_1\) & \(\alpha_2\) & \(\alpha_3\) \\
\midrule
\methane & 19.25 & \(5.213 \times 10^{-2}\) & \(1.197 \times 10^{-5}\) & \(-1.132 \times 10^{-8}\) \\
H\textsubscript{2}S & 31.94 & \(1.463 \times 10^{-3}\) & \(2.432 \times 10^{-5}\) & \(-1.176 \times 10^{-8}\) \\
n-dodecane & -9.328  & 1.149 & \(-6.347\times10^{-4}\) & \(1.359\times10^{-7}\) \\
n-tridecane & -10.46  & 1.245 & \(-6.912\times10^{-4}\) & \(1.490\times10^{-7}\) \\
n-tetradecane & -10.98  & 1.338 & \(-7.423\times10^{-4}\) & \(1.598\times10^{-7}\) \\
n-pentadecane & -11.92  & 1.433 & \(-7.972\times10^{-4}\) & \(1.720\times10^{-7}\) \\
CO\textsubscript{2} & 19.80 & \(7.344 \times 10^{-2}\) & \(-5.602 \times 10^{-5}\) & \(-1.715 \times 10^{-8}\)\\
\hydrogen  & 27.143 & \(9.274\times10^{-3}\) & \(-1.381\times10^{-5}\) & \(7.645\times10^{-9}\) \\
\nitrogen  & 31.150 & \(-1.357\times10^{-2}\) & \(2.6796\times10^{-5}\) & \(-1.168\times10^{-8}\) \\
\oxygen  & 28.106 & \(-3.680\times10^{-6}\) & \(1.7459\times10^{-5}\) & \(-1.065\times10^{-8}\) \\
\bottomrule
\end{tabular}
\end{table}

\begin{table}[H]
\centering
\caption{Parameters of Peng--Robinson EOS~\cite{smejkal_phase_2017, poling_properties_2004}.}
\label{tab:pr_params}
\begin{tabular}{l c c c}
\toprule
Component & \( T_{\text{crit}} \) [K] & \( P_{\text{crit}} \) [bar] & \(\omega\) [-] \\
\midrule
\(\methane\)    & 190.4  & 46.0  & 0.011 \\
\(\mathrm{H_2S}\)   & 373.2  & 89.4  & 0.081 \\
\hydrogen  & 33.19  & 13.00 & -0.218 \\
\nitrogen  & 126.20 & 33.98 & 0.039 \\
\oxygen  & 154.60 & 50.50 & 0.025 \\
n-dodecane   & 658.2 & 18.2 & 0.575 \\
n-tridecane  & 676.0 & 17.2 & 0.619 \\
n-tetradecane & 693.0 & 14.4 & 0.581 \\
n-pentadecane & 707.0 & 15.2 & 0.706 \\
\coo & 304.14 & 73.75 & 0.239 \\
\bottomrule
\end{tabular}
\end{table}
\clearpage

\newpage

\end{document}